\documentclass[twoside]{IEEEtran}
\usepackage[utf8]{inputenc}

\usepackage{enumerate}
\usepackage{multirow,array}
\usepackage{cite}
\usepackage{graphicx}
\usepackage{psfrag}
\usepackage{caption}
\usepackage{subcaption}
\usepackage{amsmath}
\usepackage{amsthm}
\usepackage{array}
\usepackage{amssymb}
\usepackage{amsfonts}
\usepackage{float}
\usepackage{tabu}
\usepackage{algorithm}
\usepackage{algorithmicx}
\usepackage{algcompatible}
\usepackage{algpseudocode}
\usepackage[hyphens]{url}
\usepackage{hyperref}
\usepackage{xurl}
\usepackage{blkarray}
\usepackage{bigstrut}
\usepackage{mathtools}

\hypersetup{
    colorlinks,
    citecolor=red,
    filecolor=black,
    linkcolor=blue,
    urlcolor=blue
}

\usepackage{color, colortbl}
\definecolor{Gray}{gray}{0.9}

\allowdisplaybreaks

\makeatletter
\renewcommand{\boxed}[1]{\text{\fboxsep=.2em\fbox{\m@th$\displaystyle#1$}}}
\makeatother

\newcommand{\fW}{{\tilde W}}
\newcommand{\fS}{{\tilde S}}
\newcommand{\ith}{$i^{\text{th}}$}
\newcommand{\jth}{$j^{\text{th}}$}

\newcounter{casenum}

\newcommand{\floor}[1]{\lfloor #1 \rfloor}
\newcommand{\ceil}[1]{\lceil #1 \rceil}

\newcommand{\Ladd}{L_{\mathsf{add}}}
\newcommand{\Lrem}{L_{\mathsf{rem}}}

\newtheorem*{conjecture*}{Conjecture}
\newtheorem{lemma}{Lemma}
\newtheorem{observation}{Observation}

\newtheorem{remark}{Remark}
\newtheorem{definition}{Definition}
\newtheorem{theorem}{Theorem}

\newtheorem{claim}{Claim}

\title{Coded Data Rebalancing for Distributed Data Storage Systems with Cyclic Storage}

\author{Abhinav Vaishya, Athreya Chandramouli, Srikar Kale, and
Prasad Krishnan,\IEEEmembership{ Member,~IEEE}
\thanks{
A.\ Vaishya is with Departamento de Matemática, Instituto Superior Técnico, Universidade de Lisboa, 1049-001 Lisboa, Portugal, and Instituto de Telecomunicações, 1049-001 Lisboa, Portugal. Part of this work was completed while he was with the International Institute of Information Technology, Hyderabad, India, and later with the Indian Institute of Science, Bengaluru, India (e-mail: abhinav.vaishya@tecnico.ulisboa.pt).

A. Chandramouli's work was done while at the International Institute of Information Technology, Hyderabad (e-mail: cathreya98@gmail.com). 

S. Kale is with Qualcomm, Hyderabad, India; his work was done while at the International Institute of Information Technology, Hyderabad (e-mail: srikar.kale01@gmail.com).

P.\ Krishnan is with the International Institute of Information Technology Hyderabad,
Hyderabad 500032, India (e-mail: prasad.krishnan@iiit.ac.in).

}}

\begin{document}

\maketitle

\begin{abstract}
We consider replication-based distributed storage systems in which each node stores the same quantum of data and each data bit stored has the same replication factor across the nodes. Such systems are referred to as balanced distributed databases. When existing nodes leave or new nodes are added to this system, the balanced nature of the database is lost, either due to the reduction in the replication factor, or the non-uniformity of the storage at the nodes. This triggers a rebalancing algorithm, that exchanges data between the nodes so that the balance of the database is reinstated. The goal is then to design rebalancing schemes with minimal communication load. In a recent work by Krishnan \textit{et al.}, coded transmissions were used to rebalance a carefully designed distributed database from a node removal or addition. These coded rebalancing schemes have optimal communication load, however, require the file-size to be at least exponential in the system parameters. 

In this work, we consider a cyclic balanced database (where data is cyclically placed in the system nodes) and present coded rebalancing schemes for node removal and addition in such a database. Our algorithms have time-complexity polynomial in the system parameters. These databases (and the associated rebalancing schemes) require the file-size to be only cubic in the number of nodes in the system, which we argue is the minimum required for rebalancing on such databases. We bound the advantage of our node removal rebalancing scheme over the uncoded scheme, and show that our scheme has a smaller communication load. Further, we also argue that in some parameter regimes, cyclic databases have better asymptotic tradeoffs between the rebalancing load and the file-size, when compared to earlier work by Krishnan \textit{et al}. In the node addition scenario, the rebalancing scheme presented is a simple uncoded scheme, which we show has optimal load. Finally, we show an information-theoretic lower bound for single node-removal rebalancing under the requirement of achieving a target cyclic database, and show that our achievable rebalancing loads are order-optimal (in the number of nodes), more specifically within a multiplicative gap of $2$ from the lower bound obtained. 
\end{abstract}
{\let\thefootnote\relax\footnotetext{Part of this work \cite{9965756} was presented at 2022 IEEE Information Theory Workshop held in Mumbai, India.}}

\section{Introduction}
In replication-based distributed storage systems, the available data is stored in a distributed fashion in the storage nodes with some replication factor. Doing this helps prevent data loss in case of node failures, and also provides for greater data availability and thus higher throughputs. In \cite{krishnan2020coded}, replication-based distributed storage systems in which (A) each bit is available in the same number of nodes (i.e., the replication factor of each bit is the same) and (B) each node stores the same quantum of data, were referred to as balanced distributed databases. In such databases, when a storage node fails, or when a new node is added, the `balanced' nature of the database is disturbed (i.e., the properties (A) or (B) do not hold anymore); this is known as \textit{data skew}. Data skew results in various issues in the performance of such distributed databases (see Section 1 of \cite{krishnan2020coded}). Correcting such data skew requires some communication between the nodes of the database. In distributed systems literature, this communication phase is known as \textit{data rebalancing} (see, for instance, \cite{ApacheIgniteDataRebalancing,ApacheHadoopDataRebalancing,GoogleCloud,IBMcephrebalancing}). In traditional distributed storage systems, an uncoded rebalancing phase is initiated, where uncoded bits are exchanged between the nodes to recreate the balanced conditions in the new collection of nodes. Clearly, a primary goal in such a scenario would be to minimize the communication load on the network during the rebalancing phase.

The rebalancing problem was formally introduced in the information theoretic setting in \cite{krishnan2020coded} by Krishnan \textit{et al.} The idea of \textit{coded data rebalancing} was presented in  \cite{krishnan2020coded}, based on principles similar to the landmark paper on coded caching \cite{maddah2014fundamental}. In coded data rebalancing, coded data bits are exchanged between the nodes; the decoding of the required bits can then be done by using the prior stored data available. In \cite{krishnan2020coded}, coded data rebalancing schemes were presented to rectify the data skew and reinstate the replication factor, in case of a single node removal or addition, for a carefully designed balanced database. The communication loads for these rebalancing schemes were characterized and shown to offer multiplicative benefits over the communication required for uncoded rebalancing. Information theoretic converse results for the communication loads were also presented in \cite{krishnan2020coded}, proving the optimality of the achievable loads. These results were extended to the setting of \textit{decentralized} databases in \cite{sree2021coded}, where each bit of the file is placed in some random subset of the $K$ nodes.


While Krishnan \textit{et al.} \cite{krishnan2020coded} present an optimal scheme for the coded data rebalancing problem, the centralized database design in \cite{krishnan2020coded} requires that the number of segments in the data file be a large function of the number of nodes (denoted by $K$) in the system. In fact, as $K$ grows, the number of file segments, and thus also the file size, have to grow exponentially in $K$. Thus, this scheme would warrant a high level of coordination to construct the database and perform the rebalancing. Because of these reasons, the scheme in \cite{krishnan2020coded} could be impractical in real-world settings. Motivated by this, in the present work, we study the rebalancing problem for \textit{cyclic} balanced databases, in which each segment of the data file is placed in a consecutive set of nodes, in a wrap-around fashion. For such cyclic placement, the number of segments of the file could be as small as linear in $K$. Constructing such cyclic databases and designing rebalancing schemes for them may also not require much coordination, owing to the simplicity of the cyclic placement technique.
%

We now describe the organization and contributions of this work. Section \ref{sec:cyclicdatabasemodel} sets up the formalism and gives the main result (Theorem \ref{thm:achieve}) of this work on rebalancing for single node removal or addition in a cyclic balanced database. Sections \ref{sec:rebalancing_noderemoval} and \ref{sec:rebalancing_nodeaddition} are devoted to proving this main result. In Section \ref{sec:rebalancing_noderemoval}, we present a coded data rebalancing algorithm (Algorithm \ref{algo:rebalancing}) for node removal in balanced cyclic databases. Algorithm \ref{algo:rebalancing} chooses between two coded transmission schemes (Scheme 1 and Scheme 2) based on the system parameters. Each of the two schemes has lower communication load than the other in certain parameter regimes determined by the values of $K$ and the replication factor $r$. We bound the advantage of our schemes over the uncoded scheme, and show that the minimum of their communication loads is always strictly smaller than the uncoded rebalancing scheme, which does not permit coded transmissions. Further, the segmentation required for the scheme is only quadratic in $K$, and the size of the file itself is required only to be cubic in $K$ (thus much smaller than that of \cite{krishnan2020coded}). We argue that this file-size is the minimal required, for any node-removal and node-addition rebalancing to be employed on cyclic databases.

In Section \ref{sec:rebalancing_nodeaddition}, we present a rebalancing scheme for addition of a single node to the cyclic database, and show that its load is optimal. In Section \ref{subsec:lowerboundmaintheorem_revised}, we derive a lower bound (Theorem \ref{theorem:lowr-lowerboundexpression}, \ref{theorem:highr-lowerboundexpression}) for rebalancing load for the single node removal scenario, and in Section \ref{subsec:multiGap_revised} we show that our achievable rebalancing loads are order-optimal (with respect to the number of nodes $K$), specifically within a multiplicative gap of $2$ from the lower bound (converse part of Theorem \ref{thm:achieve}). In Section \ref{sec:lowerboundcomparisons}, we show the numerical comparisons of the lower bound with the rebalancing load given in Theorem \ref{thm:achieve} and the optimal load from \cite{krishnan2020coded}.
We conclude this work in Section \ref{discussion} with several directions for future work. 

\textit{Notation:} For a non-negative integer $n$, we use $[n]$ to denote the set $\{1, \dots, n\}$. We also define $[0] \triangleq \phi$. Similarly, for a positive integer $n$, $\langle n \rangle$ denotes the set $\{0, 1, \dots, n-1\}$. To describe operations involving wrap-arounds, we define two operators. For positive integers $i, j, K$ such that $i, j \leq K$, $i \boxplus_K j = 
\begin{cases}
    i + j, & \text{ if } i+j \leq K \\
    i + j - K, & \text { if } i+j > K
\end{cases}$. Similarly, $i \boxminus_K j = 
\begin{cases}
    i - j, & \text{ if } i-j > 0 \\
    i - j + K, & \text { if } i-j \leq 0
\end{cases}$. We also extend these operations to sets. For $A \subseteq [K]$ and $i\in [K]$, we use $\{i \boxplus_K A\}$ to denote the set $\{ i \boxplus_K a \; : \; a \in A \}$. Similarly, $\{i \boxminus_K A\}$ denotes $\{ i \boxminus_K a \; : \; a \in A \}$. For a binary vector $X$, we use $|X|$ to denote its size. The concatenation of two binary vectors, $X_1$ and $X_2$, is denoted by $X_1|X_2$. In a directed graph $\cal G$, the \textit{subgraph induced by a subset $A$ of its vertices} is the graph with $A$ as its vertices and all the edges in $\cal G$ between vertices in $A$ as its edges.  

\subsection{Relationship to Other Data Distribution Problems with Side-Information}
\label{subsec:Relationship}

Index Coding, introduced by Birk and Kol \cite{ISCOD_BiK}, and Network Coding \cite{ahslwedecailiyeung_NIF_2000} introduced by Ahlswede, Cai, Li and Yeung demonstrated the potential of  `information flows' (loosely, encoding multiple data streams meant for distinct users into one) to increase the rate of data delivery to multiple clients on a common network. In network coding, multiple encoded streams are decoded together by any client, to obtain their desired data. In index coding, multiple encoded transmissions are decoded by clients with the assistance of existing \textit{side-information} (message symbols pre-stored in local memory), to obtain their specific desired symbols. A slew of communication problems were studied in the subsequent years, which applied ideas from network coding and index coding to data distribution on a network, while exploiting side-information whenever available. We first summarize a few such works that are closely related to ours, and highlight the distinctive aspects of this work at the end of this subsection. 

A major research area in which the paradigm of network coding was  applied to obtain novel results and constructions was erasure codes for distributed storage \cite{dimakis_etal_NC_distributedstorage_2010,dimakis_IEEEproceeddings_NCforDistributedStorage,LRC_2014}. Further, the  work \cite{maddah2014fundamental} by Maddah-Ali and Niesen utilized the index coding paradigm to show that considerable coding gains can be obtained in broadcast networks for delivery of files simultaneously to multiple clients, when employing a carefully designed pre-fetching (or, caching) phase that populated caches available at the clients. This new framework, known as \textit{coded caching}, was then extended to other problems in distributed communication and computation, such as distributed linear computation \cite{Lee_etal_2018_speedingupML} and distributed data-shuffling in Map-Reduce frameworks \cite{LiAliYu_TIT_FundamentalTradeoffCompComm_MapReduceOriginalJournal_2018}. 

The central works \cite{maddah2014fundamental,Lee_etal_2018_speedingupML,LiAliYu_TIT_FundamentalTradeoffCompComm_MapReduceOriginalJournal_2018} and subsequent literature in these areas typically considered a single master-node with multiple clients/workers setup, in which the master was connected to the clients via a broadcast channel. The cache content available at the client nodes enables multiple coding opportunities during the demand phase, when the data has to be delivered to the clients. Thus, by appropriate design of the cache-placement and the delivery scheme, remarkable multiplicative coding gains can be obtained in the total communication load in the delivery phase. Enabling this careful design of the caching phase often requires a large \textit{subpacketization} (effectively file-size), which implies increased implementation complexity \cite{user_grouping_Shanmugam}. Variety of low-subpacketization schemes have been considered in the area of coded caching and distributed computing, via a number of combinatorial and coding-theoretic techniques (for instance, see \cite{user_grouping_Shanmugam,strongedgecoloringofbipartitegraphs,TaR,hari_TIT_2021_subexp}). 

With the aim of reducing subpacketization, cyclic placement schemes, in which the files are placed in the caches in a cyclic manner, have also been considered before, in the context of distributed storage \cite{CyclicStorage_Distirbuted_TPDS, Cyclic_Replication_Distributed}, multi-antenna wireless coded caching \cite{salehietal_2022_TWC_cyclicplacement_MISO,MultiantennaPDA_cyclic_Kaietal_2022}, as well as coded data shuffling and distributed computing \cite{KaiWan_Sun_etal_2022_TIT_DistributedLinearComp_cyclicplacement,pmlrv70tandon17a_gradientcoding,Cyclic_Storage_DistributedComputing_Mingyue,wan2021tradeoff,huang2023fundamental,huang2025optimality}. Decentralized setups \cite{aliniesen_2015_decentralizedACMnetworking,kaiwan_etal_2018_Allerton_distributeddecentralizeddatashuffling} have also been considered in coded caching and data shuffling, in which data is exchanged between the client/worker nodes without the presence of the master node. In distributed computation, we also have the issue of \textit{straggler} nodes, which are nodes that are slow in performing the assigned computational task. Clever file-placement strategies and coding-theory-based delivery schemes have been proposed to overcome the effect of straggler nodes, in a number of works \cite{KaiWan_Sun_etal_2022_TIT_DistributedLinearComp_cyclicplacement,pmlrv70tandon17a_gradientcoding,Cyclic_Storage_DistributedComputing_Mingyue,wan2021tradeoff,huang2023fundamental}.

In the context of coded distributed storage, data is encoded using a block code (to protect against node failures) and these encoded symbols stored in a distributed fashion across multiple nodes, which can also perform computations and have communication capabilities. When one of these storage nodes fail, it becomes necessary to reconstruct the encoded data that was present in these nodes, in a new empty node that enters the system. This repair process is done by transmissions of symbols (possibly after encoding) from the existing nodes, to the new node, which will then run a decoding process. The network coding paradigm is useful in the process of designing these transmissions. In \textit{exact repair}, we wish to make the new node an identical copy of the failed node. Thus, the distributed storage code that encodes the data must serve two purposes: (a) to protect against node-failures, and (b) to reduce the overall communication required to perform the node-repair, after any failure. Such codes are thus called \textit{exact-repair regenerating codes} \cite{dimakis_etal_NC_distributedstorage_2010,exact_regeneratingRashmiNihar_2011,salim_etal_2010_fractional_allerton}.

Another related line of work involves coded data shuffling in distributed learning environments, which was initiated in \cite{Lee_etal_2018_speedingupML,LiAliYu_TIT_FundamentalTradeoffCompComm_MapReduceOriginalJournal_2018}. The general version of this setting, as presented in subsequent works, \cite{attiatandon_TIT_2019,elmahdy2020_TIT_limitsofCodedDataShufflingDistributeDlearning} considers a distributed learning environment with one master and multiple workers. The available data is divided into files and distributed by the master node, across a collection of worker nodes. Each worker node can store only a fraction of the complete data (some number of files) in its local storage (or cache). The workers aid the master in computing some global function of the data, by performing and returning some local functions on their assigned files, over multiple iterations. Between iterations, the files available in each node are to be altered (\textit{shuffled}), according to some global permutation. To effect this change, the master sends coded transmissions to all the workers via a broadcast channel. This phase, known as  \textit{data-shuffling}, therefore works as per the index coding paradigm, as the master and workers exploit the knowledge of cached data in the prior iteration, to construct coded transmissions and decode the required data in the subsequent iteration respectively. In this process, the initial structure of the distributed storage is typically  maintained in each iteration. For instance, the works \cite{attiatandon_TIT_2019,elmahdy2020_TIT_limitsofCodedDataShufflingDistributeDlearning} consider uncoded data placement with some replication structure, which has to be maintained.

The data-rebalancing problem on a distributed storage system (called in the present work a distributed database), first considered in \cite{krishnan2020coded} and continued in this present work, considers another setting that requires data delivery in the presence of side-information. In the present work, we consider a master-free network of nodes, which contain the data in a cyclic fashion, similar to \cite{Cyclic_Replication_Distributed,Cyclic_Storage_DistributedComputing_Mingyue,CyclicStorage_Distirbuted_TPDS,huang2023fundamental} 
. While there is similarity in the techniques developed for data delivery in coded caching or the distributed data shuffling scenario with cyclic placement, the requirement of \textit{rebalancing} is specific to the present work. A distinct aspect in rebalancing is that, the demands for the nodes during rebalancing is to be fixed carefully prior to the design of the scheme itself, so as to minimize the rebalancing load. Indeed, the quantum of data delivered to the nodes during rebalancing also need not be uniform, in our work. In this sense, the rebalancing problem differs from coded caching and data shuffling, both in which node-demands are chosen from a predefined collection of types. 

Further, the data-rebalancing problem considered in \cite{krishnan2020coded} and in this work is similar to the node-repair problem \cite{dimakis_etal_NC_distributedstorage_2010,exact_regeneratingRashmiNihar_2011,salim_etal_2010_fractional_allerton}, in the sense of being able to recover from node-removals, by maintaining an identical redundancy-structure in the target database. However, in the rebalancing setting, a node-removal does not trigger an addition of a new empty node. Instead, the existing nodes must bear the cost of storing more data, so that the overall replication factor (i.e., redundancy level) remains the same for the entire data.  

The achievability schemes we present are three-phased: (i) choosing the data to be delivered to the nodes, (ii) design of the index-coding based scheme to deliver the data, and (iii) merging the delivered data together, to obtain the new database. Further, we present the converse arguments which lower-bounds the minimum amount of data to be transmitted to complete the rebalancing, provided the final database is also having the cyclic placement structure. The rebalancing load achieved by our scheme is within a factor of $2$ from the converse bounds we obtain. 





    
\section{Rebalancing Schemes for Cyclic Databases}
\label{sec:cyclicdatabasemodel}
In this section, we give the formal definition of cyclic databases and present our main result (Theorem \ref{thm:achieve}) on rebalancing schemes for node removal and addition for such databases. Towards that end, we recall the formal system model and other relevant definitions from \cite{krishnan2020coded}. 

Consider a binary file $W$ consisting of a set of $N$ equal-sized segments where the \ith segment is denoted by $W_i$, where $|W_i|=T$ bits. The system consists of $K$ nodes indexed by $[K]$ and each node $k \in [K]$ is connected to every other node in $[K] \backslash \{k\}$ via a bus link that allows a noise-free broadcast between the nodes. A \textit{distributed database} of $W$ across nodes indexed by $[K]$ is a collection of subsets $\mathcal{D}= \{D_k \subseteq \{W_i:i\in[N]\} \; : \;  k \in [K] \}$, such that $\bigcup_{k\in[K]} D_k = \{W_i:i\in[N]\}$, where $D_k$ denotes the set of segments stored at node $k$. We will denote the set of nodes where $W_i$ is stored as $S_i$. The \textit{replication factor} of the segment $W_i$ is then $|S_i|$. The distributed database $\cal D$ is said to be \textit{$r$-balanced}, if $|S_i|=r, \forall i$, and $|D_k|=\frac{rN}{K}, \forall k$. That is, each segment is stored in $r$ nodes, and each node stores an equal $\frac{r}{K}$-fraction of the $N$ segments. We may assume without loss of generality that $2\leq r\leq K-1$, since if $r=1$ no rebalancing is possible from node removal, and if $r=K$ no rebalancing is required. 

When a node $k$ is removed from the $r$-balanced database, the replication factor of the segments $D_k$ stored in node $k$ drops by one, thus disturbing the `balanced' state of the database. If a new empty node $K+1$ is added to the database, once again, the new database is not balanced. To reinstate the balanced state, a \textit{rebalancing scheme} is initiated, which is formally defined as follows.

\begin{definition}[Rebalancing Scheme]
\label{defn:rebalancing}
A \textit{rebalancing scheme} $\mathcal{S}$ for the removal of node $k$ 
from a database $\mathcal{D}$ on nodes $[K]$ consists of:
\begin{enumerate}
    \item A target $r$-balanced database $\mathcal{D}' = \{D'_{k_1} : 
    k_1 \in [K]\setminus\{k\}\}$ on nodes $[K]\setminus\{k\}$.
    \item For each node $k_1 \in [K]\setminus\{k\}$, an encoding function 
    $\phi_{k_1}$ that maps the 
    local storage $D_{k_1}$ to a transmitted message denoted as $X_{k_1} = \phi_{k_1}(D_{k_1})$.
    \item For each node $k_1 \in [K]\setminus\{k\}$, a decoding function 
    that uses $D_{k_1}$ and $\{X_{k_2}: k_2\in [K]\setminus\{k\}\}$ to 
    reconstruct $D'_{k_1}$.
\end{enumerate}
A scheme $\mathcal{S}$ is said to be \textit{feasible} if the decoding 
succeeds at every node $k_1\in[K]\setminus\{k\}$, i.e., $D'_{k_1}$ is 
correctly recovered at node $k_1$ for all $k_1\in[K]\setminus\{k\}$. The 
\textit{demand} $B_{k_1}$ at node $k_1$ under scheme $\mathcal{S}$ is 
defined as $B_{k_1} \triangleq D'_{k_1}\setminus D_{k_1}$, i.e., the bits 
required at node $k_1$ in the target database that were not present 
prior to rebalancing. The communication load of scheme $\mathcal{S}$ is 
\begin{align}
\label{eqn:defnsofrebalancingloads}
   L_{\mathsf{rem},{\cal S}}(r)=\frac{\sum_{k_1\in[K]\setminus\{k\}}{|X_{k_1}|}}{T}.
\end{align}
Similar definitions apply for the node addition case, with target database 
$\mathcal{D}'$ on nodes $[K+1]$, transmissions from all $K$ existing nodes, 
and load $L_{\mathsf{add},\mathcal{S}}(r) = \frac{\sum_{k_1\in[K]}|X_{k_1}|}{T}$.
\end{definition}
Now, it is easy to see that the rebalancing load for node removal using the uncoded scheme is $L_u(r)=r$. The \textit{normalized} communication loads are then defined as $\mathfrak{L}_{rem,{\cal S}}(r)=L_{\mathsf{rem},{\cal S}}(r)/N$ and $\mathfrak{L}_{add,{\cal S}}(r)=L_{\mathsf{add},{\cal S}}(r)/N$. The \textit{optimal} normalized communication loads for the node removal 
and addition scenarios are denoted by $\mathfrak{L}_{rem}^*(r)$ and 
$\mathfrak{L}_{add}^*(r)$ respectively, where the optimality is by 
minimization over all choices of $N$, $T$, initial and target $r$-balanced 
databases, and all feasible rebalancing schemes $\mathcal{S}$.

\begin{figure}[t]
\includegraphics[scale=0.2]{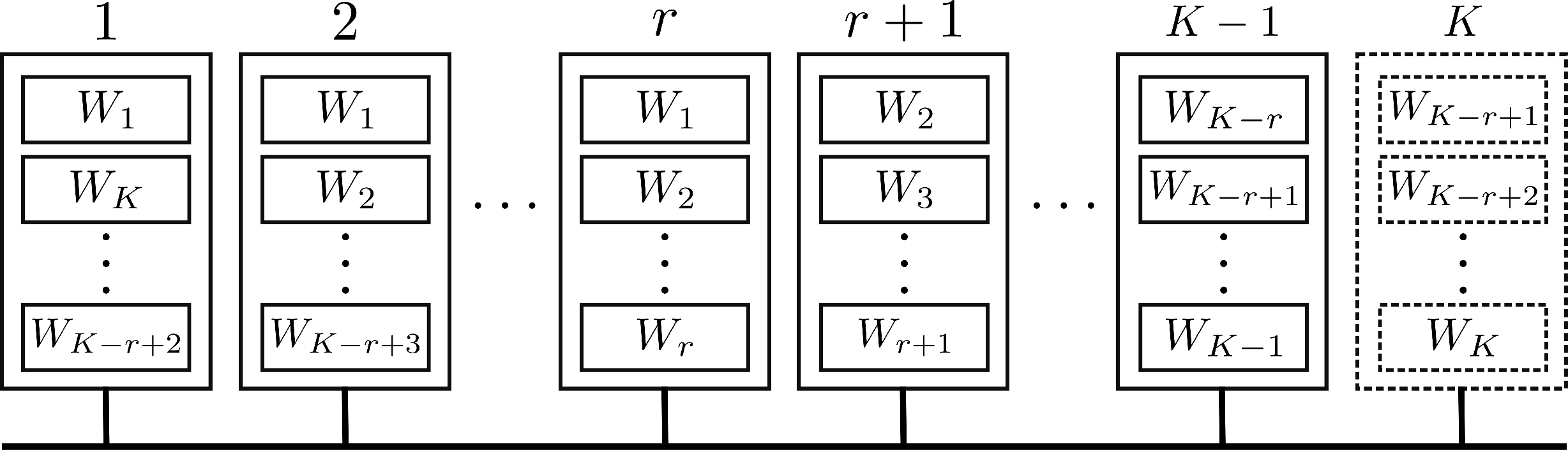}
\centering
\caption{$r$-balanced cyclic database on nodes $[K]$}
\label{fig:cycliconKnodes}
\end{figure}

In \cite{krishnan2020coded}, it was shown\footnote{The size of the file in the present work is $NT$ bits; whereas in \cite{krishnan2020coded}, the notation $N$ represents the file size in bits, thus absorbing both the segmentation and the size of each segment. The definitions of communication loads in \cite{krishnan2020coded} are also slightly different, involving a normalization by the storage size of the removed (or added) node. The results of \cite{krishnan2020coded} are presented here according to our current notations and definitions.} that $\mathfrak{L}_{rem}^*(r)\geq \frac{r}{K(r-1)}$ and $\mathfrak{L}_{add}^*(r)\geq \frac{r}{K+1}$. Further, schemes for rebalancing were presented for node removal and addition for a carefully designed database which required $N=\frac{(K+1)!}{r!}$ and $T=r-1$, which achieve these optimal loads. Observe that, in these achievable schemes, the file size $NT$ grows (at least) exponentially in $K$ as $K$ grows, for any fixed replication factor $r$, which is one of the main drawbacks of this result. Therefore, our interest lies in databases where $N$ and $T$ are small. Towards this end, we now define \textit{cyclic} databases. 
\begin{definition}
\label{defn:cyclic}
A distributed database is an $r$-balanced cyclic database if $N = K$ and a segment labelled $W_i$ is stored precisely in the nodes $S_i=\{ i \boxplus_K \langle r \rangle \}$. In other words, in an $r$-balanced cyclic database, the set of segments placed in node $k$ is precisely $D_k=\{W_i:i\in \{k\boxminus_K \langle r\rangle\}\}$.
\end{definition}
 Fig. \ref{fig:cycliconKnodes} depicts an $r$-balanced cyclic balanced database on $K$ nodes as defined above. In this work, we present rebalancing schemes for node removal and addition, for such a cyclic database on $K$ nodes. Specifically, we prove the following result. 	

\begin{theorem}
\label{thm:achieve}
For an $r$-balanced cyclic database having $K$ nodes and $r\in \{3,\hdots,K-1\}$, if the segment size $T$ is divisible by $2(K^2-1)$, then rebalancing schemes for node removal and addition exist which achieve the respective communication loads
\begin{align}
\label{eqn:rebalancingload_noderemoval}
\Lrem(r) & = \frac{K-r}{(K-1)}+\min\left(L_1(r), L_2(r) \right),\\
\label{eqn:rebalancingload_nodeaddition}
\Ladd(r) &= \frac{rK}{K+1},
\end{align}
where, $L_1(r) = \frac{(K-r)(2r-1)}{(K-1)}$ 
and $L_2(r) = \frac{1}{2(K-1)}\left(K(r-1)+ \ceil{\frac{r^2-2r}{2}}\right)$. Also, the following statements hold true. 
\begin{enumerate}
    \item  Let $L_u(r)$ denote the load of uncoded rebalancing scheme for node removal. Then, 
 {\small
\begin{align*}
\frac{\Lrem(r)}{L_u(r)}
&<
\min\Bigg(
2\left(
1-\frac{r-1}{K-1}
\right),
\nonumber \\
&\qquad
\left(
\frac{1}{2}
+
\frac{1}{2r}
+
\frac{r}{4(K-1)}
\right)
\Bigg)
<1.
\end{align*}
}
\item Let $\Lrem^*(r)$ denote the optimal (smallest) load among all rebalancing schemes that achieve a cyclic $r$-balanced target database on $K-1$ nodes, for the node removal scenario. Then, 
 \begin{align*}\frac{\Lrem(r)}{\Lrem^*(r)} \leq \begin{cases}
 \frac{5}{4}, & \text{if}~r< \frac{K+1}{2}.\\
 \frac{4}{3}, & \text{if}~r = \frac{K+1}{2}.\\
 2, & \text{if}~\frac{K+1}{2}<r\leq K-1.
 \end{cases}
 \end{align*}
 \item The rebalancing scheme for node addition is optimal (i.e.,  $\Ladd(r)/N=\mathfrak{L}_{add}^*(r)=\frac{r}{K+1}$).
\end{enumerate}
\end{theorem}

\begin{remark}
In the proof of the node removal part of Theorem \ref{thm:achieve}, we assume that the target database is also cyclic. For the case of $r=2$, with this target database, our rebalancing scheme does not apply, as coding opportunities do not arise. Hence, we restrict our result to the scenario of $r\in\{3,\hdots,K-1\}$.
\end{remark}
\begin{remark}[Asymptotic behaviour of $\Lrem(r)$ and asymptotic multiplicative gap from $\Lrem(r)$]
\label{rem:asymptoticgap}
Observe that, for constant $r$, as the number of nodes $K$ grows asymptotically, we get $\Lrem(r)=\Theta(1)$, showing the same asymptotic behaviour as the optimal load $\frac{r}{r-1}$ (achieved with large file-size) from \cite{krishnan2020coded}. Further, interestingly, if $r=K-\rho$, for some constant $\rho$, then $L_1(r)=\Theta(1)$ and $L_2(r)=\Theta(K)$, as $K$ grows large. Thus, in this regime too, we observe that $\Lrem(r)$ has the same asymptotic behaviour as the optimal load, i.e., $\Lrem(r)=\Theta(1)$. Thus, in these two regimes, opting for cyclic placement might be a reasonable choice, as the tradeoff between file-size and the load can be advantageous for cyclic placement, than for the scheme in \cite{krishnan2020coded}, which can require subpacketization at least $\Theta(K^{\min(K-r,r)})$ for this regime.

However, if both $r$ and $K-r$ grow as $\Theta(K)$, then gap between the optimal load from \cite{krishnan2020coded} and $\Lrem(r)$ widens. Indeed, it is not difficult to see from \eqref{eqn:rebalancingload_noderemoval} that, in this regime, we have $\Lrem(r)=\Theta(K)$ as both $L_1(r)$ and $L_2(r)$ behave as $\Theta(K)$. However, in this case, the optimal load still is $\Theta(1)$. Thus, in this `middle' regime, the gap between the optimal load $\frac{r}{r-1}$ and $\Lrem(r)$ widens as $\Theta(K)$. These aspects regarding the gap between the optimal load from \cite{krishnan2020coded} and the present achievable load are also observed in Fig. \ref{fig:lowerbound} in Section \ref{sec:lowerboundcomparisons}, which compares our achievable load with the lower bound in the present work as well as the optimal load from \cite{krishnan2020coded}.

  Now we comment on the asymptotic behaviour of the multiplicative gap between the optimal load for a cyclic target database $\Lrem^*(r)$ and the achievable load $\Lrem(r)$. If $r$ is kept constant, the multiplicative gap between the optimal node-removal rebalancing load $\Lrem^*(r)$ and the achievable load becomes at most $\frac{5}{4}$, as the regime $r< (K+1)/2$ becomes applicable for large enough $K$. However, if both $r$ and $K$ grow simultaneously, then the  multiplicative gap of $2$ is still applicable as it holds for each $r$ and $K$. Thus, there is no marked asymptotic increase in the loss of optimality as $K$ grows, as long as we seek to achieve a target cyclic database. 
\end{remark}
Theorem \ref{thm:achieve} is proved via Sections \ref{sec:rebalancing_noderemoval},  \ref{sec:rebalancing_nodeaddition} and \ref{sec:lower_bound_revised}. In Section \ref{sec:rebalancing_noderemoval}, we prove the achievability results in Theorem \ref{thm:achieve} regarding the node removal scenario. Specifically, we prove the expression (\ref{eqn:rebalancingload_noderemoval}) and the comparison with the uncoded load, as in part \textit{1)}. We present a rebalancing algorithm, the core of which is a transmission phase in which coded subsegments are communicated between the nodes. In the transmission phase, the algorithm chooses between two schemes, Scheme 1 and Scheme 2. Scheme 1 has the communication load $\frac{K-r}{K-1}+L_1(r)$ and Scheme 2 has the load $\frac{K-r}{K-1}+L_2(r)$. We identify a threshold value for $r$, denoted $r_{\mathsf{th}}$, beyond which Scheme 1 is found to be performing better than Scheme 2, as shown by the following claim; and thus the rebalancing algorithm chooses between the two schemes based on whether $r\geq r_{\mathsf{th}}$ or otherwise. The proof of the below claim is in Appendix \ref{appendix proof threshold}.
\begin{claim}
\label{claim:threshold}
Let $r_{\mathsf{th}}=\ceil{\frac{2K+2}{3}}$. If $r \geq r_{\mathsf{th}}$, then $\min(L_1(r),L_2(r))=L_1(r)$ (thus, Scheme 1 has a smaller load) and if $r<r_{\mathsf{th}}$, we have  $\min(L_1(r),L_2(r))=L_2(r)$ (thus, Scheme 2 has a lower communication load). 
\end{claim}

\begin{figure}[h]
\includegraphics[scale=0.6]{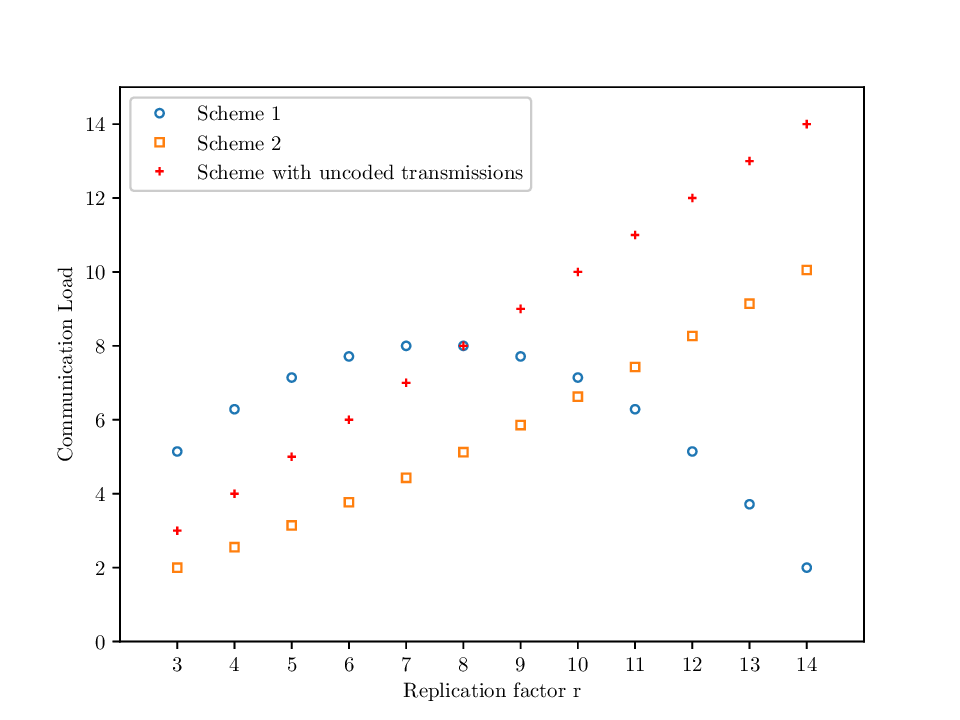}
\centering
\caption{\small For $K=15$, the figure shows comparisons of communication loads of Scheme 1 and Scheme 2 with the load of uncoded transmission scheme, for varying $r$. Note that all curves are relevant only for $r\in\{3,\hdots,K-1\}$. We see that the minimum of the loads of the two schemes is always less than the uncoded load. Further, by Claim \ref{claim:threshold}, any integer value of $r\geq 11$, we see that Scheme 1 has smaller load than Scheme 2, and the reverse is true otherwise.} 
\label{fig:communicationLoad}
\end{figure}

A comparison of these schemes is shown in Fig. \ref{fig:communicationLoad} for the case of $K=15$ (as $r$ varies), along with the load $L_u(r)$ of the uncoded rebalancing scheme.

In Section \ref{sec:rebalancing_nodeaddition}, we show a rebalancing scheme for node addition. We prove that the communication load of this scheme is given by \eqref{eqn:rebalancingload_nodeaddition}, and show its optimality, i.e., prove \textit{part 3)} of Theorem \ref{thm:achieve}. In Section \ref{sec:lower_bound_revised}, we derive lower bounds (for various regimes of $r,K$) for the node-removal rebalancing load, and show that these are tight upto the multiplicative constants, as given in part \textit{2)} of Theorem \ref{thm:achieve}.

We observe that the minimum file size required for our achievability scheme as per Theorem \ref{thm:achieve} is $NT=2(K^2-1)K$, i.e., it is cubic in $K$ and thus much smaller than the file size requirement for the schemes of \cite{krishnan2020coded}. Further, the following observation suggests that this is infact order-wise optimal, if the target database is cyclic, as we observe below. 

\begin{observation}[Regarding Optimality of File-size in Theorem \ref{thm:achieve}]
In any $r$-cyclic database, to facilitate rebalancing due to a node-removal, we must have $(K-1)|T$, in general. This is trivially true, since each of the $K-1$ new segments in the cyclic $r$-balanced target database must be of size $\frac{KT}{K-1}$, where $T$ is the size of each of the $K$ segments in the initial cyclic database. Thus, in general, $(K-1)$ must divide $T$, hence giving our result. Further, by a similar argument, for node-addition, we see that $(K+1)|T$. Thus, assuming that there exist infinite sequences of twin-primes \cite{Maynard2019}, the segment-size for any rebalancing scheme that incorporates both node-removal and node-addition rebalancing must be $\Theta(K^2)$. Including the fact that the number of segments must be $K$ for a cyclic database, we see that the file-size must be $\Theta(K^3)$. Thus, the file-size used in the rebalancing schemes of Theorem \ref{thm:achieve} is order-wise optimal (as a function of $K$). 
\end{observation}

\section{Proof of Theorem \ref{thm:achieve}: Rebalancing for Single Node Removal in Cyclic Databases}
\label{sec:rebalancing_noderemoval}
In Subsection \ref{sub:intuition}, we provide some intuition for the rebalancing algorithm, which covers both transmission schemes, Scheme 1 and Scheme 2. Then, in Subsection \ref{sub:example}, we describe how the algorithm works for two example parameters (one for Scheme 1, and another for Scheme 2). In Subsection \ref{sub:algorithm}, we formally describe the complete details of the rebalancing algorithm. In Subsection \ref{sub:correctness}, we prove the correctness of two transmission schemes and the rebalancing algorithm. In Subsection \ref{subsec:comm_load_noderemoval}, we calculate the communication loads of our schemes. Finally, in Subsection \ref{subsec:advantage}, we bound the advantage of our schemes over the uncoded scheme, and show that our schemes perform strictly better, thus completing the arguments for the node-removal part of Theorem \ref{thm:achieve}.

\begin{remark}Note that throughout this section, we describe the scheme when the node $K$ is removed from the system. A scheme for the removal of a general node can be extrapolated easily by permuting the labels of the subsegments. Further details are provided in Subsection \ref{sub:algorithm} (see Remark \ref{remark:arbitrarynoderemoval}).
\end{remark}

\subsection{Intuition for the Rebalancing Scheme}
\label{sub:intuition}
\begin{figure}[h]
\includegraphics[scale=0.2]{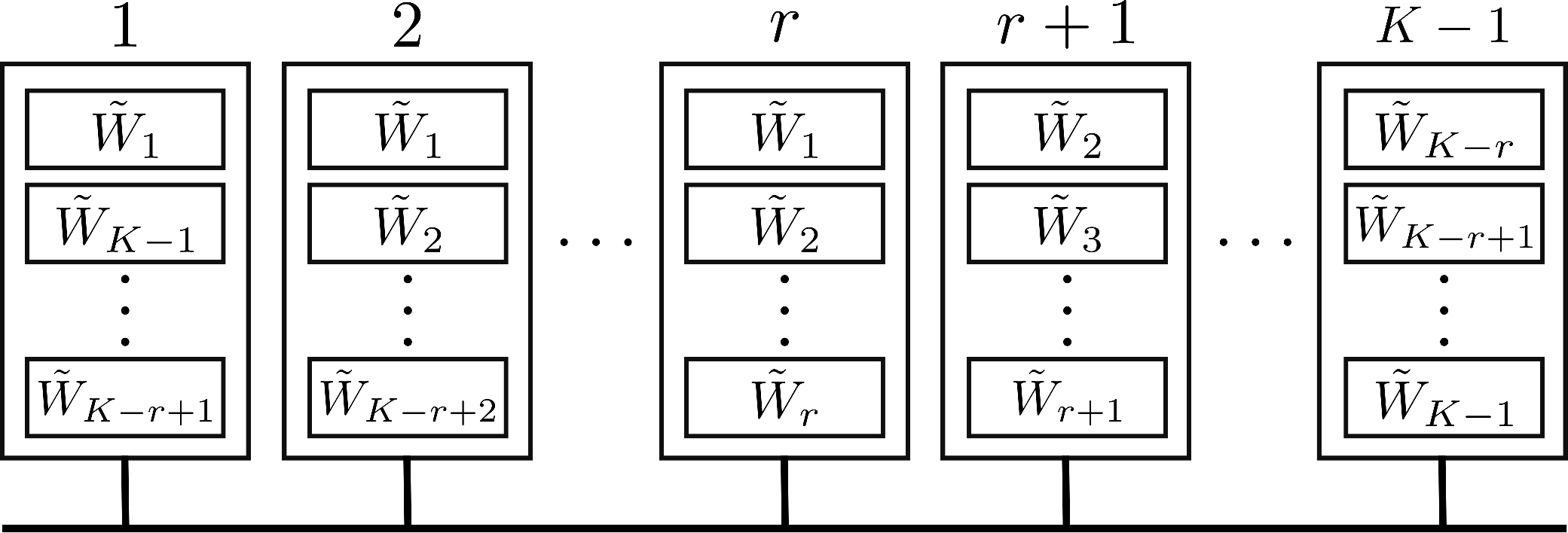}
\centering
\caption{Target cyclic database on nodes $[K-1]$}
\label{fig:cycliconK-1nodes}
\end{figure}

Consider an $r$-balanced cyclic database as shown in Fig. \ref{fig:cycliconKnodes}. Without loss of generality, consider that the node $K$ is removed. Now the segments that were present in node $K$, i.e., $D_K=\{W_{K-r+1}, \dots, W_{K}\}$, no longer have replication factor $r$. In order to restore the replication factor of these segments, we must reinstate each bit in these segments via rebalancing into a node where it was not present before. We fix the target database post-rebalancing to also be an $r$-balanced cyclic database. Recall that $S_i=\{i\boxplus{_{K}}\langle r\rangle\}$ represents the nodes where $W_i$ was placed in the initial database. We represent the $K-1$ file segments in this target cyclic database as $\fW_i:i\in[K-1]$, and the nodes where $\fW_i$ would be placed is denoted as $\fS_i=\{i\boxplus{_{K-1}}\langle r\rangle\}$. This target database is depicted in Fig. \ref{fig:cycliconK-1nodes}. 

Our rebalancing algorithm involves three phases: (a) a \textit{splitting phase} where the segments in $D_K$ are split into subsegments, (b) a \textit{transmission phase} in which coded subsegments are transmitted, and (c) a \textit{merge phase}, where the decoded subsegments are merged with existing segments, and appropriate deletions are carried out, to create the target database. Further, the algorithm will choose one of two transmission schemes, Scheme 1 and Scheme 2, in the transmission phase. Both of these schemes are XOR-coding based schemes, and essentially employ the well-known `clique-cover' technique in index coding \cite{ISCOD_BiK}. Our discussion here pertains to both these schemes. 

The design of the rebalancing algorithm is driven by two natural objectives: (a) move the subsegments as minimally as possible, and (b) exploit the available coding opportunities. Based on this, we give the three generic principles below. 
\begin{itemize}
    \item \textit{Principle 1:} The splitting and merging phases are unavoidable for maintaining the balanced nature of the target database and reducing the communication load. In our merging phase, the target segment $\fW_j:j\in[K-1]$ is constructed by merging, (a) either a subsegment of $W_j$  or the complete $W_j$, along with  (b) some other subsegments of the segments in $D_K$. 
    \item \textit{Principle 2:} Particularly, for each $W_i\in D_K$, we seek to split $W_i$ into subsegments and merge these into those $\fW_j:j\in[K-1]$ such that $|\fS_j\cap S_i|$ is as large as possible, while trying to ensure the balanced condition of the target database. Observe that the maximum cardinality of such intersection is $r-1$.  We denote the subsegment of segment $W_i$ which is to be merged into $\fW_j$, and thus to be placed in the nodes $\fS_j\setminus S_i$, as $W_i^{\fS_j\setminus S_i}$. As  making $|\fS_j\cap S_i|$ large reduces  $|\fS_j\setminus S_i|$, we see that this principle reduces the movement of subsegments during rebalancing. 
    \item \textit{Principle 3:}  Because of the structure of the cyclic placement, there exist `nice' subsets of nodes whose indices are separated (cyclically) by $K-r$, which provide coding opportunity. In other words, there is a set of subsegments of segments in $D_K$, each of which is present in all-but-one of the nodes in any such `nice' subset, and is to be delivered to the remaining node. Transmitting the XOR-coding of these subsegments ensures successful decoding at the respective nodes they are set to be delivered to (given by the subsegments' superscripts), because of this `nice' structure. These subsets which provide coding opportunities were identified by careful observation of the relationships between the demanded subsegments and the content already available in the nodes.
\end{itemize}

We shall illustrate the third principle, which guides the design of our transmission schemes, via the examples and the algorithm itself. We now elaborate on how the first two principles are reflected in our algorithms. Consider the segment $W_{K-r+1}$. We call this a \textit{corner} segment of the removed node $K$.  Following Principles 1 and 2, this segment $W_{K-r+1}$ will be split into $\ceil{\frac{K-r+2}{2}}$ subsegments, out of which one large subsegment is to be merged into $\fW_{K-r+1}$ (as $|S_{K-r+1}\cap\fS_{K-r+1}|=r-1$). Observe that this subsegment is already available in nodes $\{K-r+1,\hdots,K-1\}$ and needs to be therefore moved only to node $1$, as $\fS_{K-r+1}\setminus S_{K-r+1}=\{K-r+1,\hdots,K-1,1\}\setminus \{K-r+1,\hdots,K\}=\{1\}$. 

The remaining $\ceil{\frac{K-r+2}{2}}-1$ subsegments of $W_{K-r+1}$ are to be merged into the $\floor{\frac{K-r}{2}}$ target segments $\{\fW_{\ceil{\frac{K-r}{2}}+1}, \hdots,\allowbreak 
\fW_{K-r}\}=\{\fW_{K-r-j+1}:j\in [(K-r)-\ceil{\frac{K-r}{2}}]\}$, and additionally into the segment $\fW_{\frac{K-r+1}{2}}$ if $K-r$ is odd. Again, the reason for merging parts of $W_{K-r+1}$ into $\fW_{\ceil{\frac{K-r}{2}}+1}, \hdots, \fW_{K-r}$ is because \textit{(a)} we want a $r$-balanced cyclic target database, and \textit{(b)} among all the location-subsets of the target segments $\fW_j:j\in[K-1]$ (apart from $\fW_{K-r+1}$ which has already been considered), the intersection-sizes $|S_{K-r+1}\cap\fS_{K-r-j+1}|=\max(r-j,0)$, for $j\in \{1,\hdots,(K-r)-\ceil{\frac{K-r}{2}}\}$, are the largest. This implies that the subsegment to be merged into $\fW_{K-r-j+1}$ has to be moved to precisely $\min(r,j)$ nodes, which are $\fS_{K-r-j+1}\setminus S_{K-r+1}=\{K-r-j+1,\hdots,(K-r-j)+\min(r,j)\}$. Similar arguments can be written for the subsegment to be merged with $\fW_{\frac{K-r+1}{2}}$ as well (when $(K-r)$ is odd).  

The other \textit{corner} segment of $K$ is $W_K$, for which a similar splitting is followed. One large subsegment of  $W_K$ will be merged into $\fW_{K-1}$ (again, as $|S_{K} \cap \fS_{K-1}| = r-1$), which is to be delivered to $\fS_{K-1}\setminus S_K=\{K-1\}$  . The remaining $\ceil{\frac{K-r+2}{2}}-1$ subsegments will be merged into $\floor{\frac{K-r}{2}}$ target segments $\fW_1, \hdots, \fW_{\floor{\frac{K-r}{2}}}$, and additionally into the segment $\fW_{(K-r+1)/2}$ if $K-r$ is odd. Specifically, the target nodes for the subsegment to be merged with $\fW_j:j\in[\floor{\frac{K-r}{2}}]$ is to be delivered to the $\min(r,j)$ nodes $\fS_j\setminus S_K=\{(r+j-1),(r+j-1)-1,\hdots,r+j-\min(r,j)\}$. Similar arguments can be written for the subsegment to be merged with $\fW_{(K-r+1)/2}$ as well, when $K-r$ is odd.

Now, consider the segment $W_{K-r+2}$. This is not a corner segment, hence we refer to this as a \textit{middle} segment. This was available in the nodes $S_{K-r+2}=\{K-r+2,\hdots,K,1\}$. Following Principles 1 and 2, this segment $W_{K-r+2}$ will split into two: one to be merged into $\fW_{K-r+1}$ (for which $\fS_{K-r+1}=\{K-r+1,\hdots,K-1,1\}$, and thus to be placed in $\{K-r+1\}$) and $\fW_{K-r+2}$ (for which $\fS_{K-r+2}=\{K-r+2,\hdots,K-1,1,2\}$ and thus to be placed in $\{2\}$). Observe that $|S_{K-r+2}\cap\fS_{K-r+1}|=r-1$ and $|S_{K-r+2}\cap\fS_{K-r+2}|=r-1$. In the same way, each middle segment $W_{K-r+i+1}:i\in[r-2]$ is split into two subsegments, each one respectively to be placed in nodes $K-r+i$ and $i+1$, and to be merged into $\fW_{K-r+i}$ and $\fW_{K-r+i+1}$ respectively. 

The sizes of these subsegments described above are given precisely in splitting algorithm (Algorithm \ref{algo:split}) in Subsection \ref{sub:algorithm}, illustrated via Fig. \ref{fig:split1}, \ref{fig:split2}, and \ref{fig:split3}. The intuition behind fixing the sizes of the subsegments, are discussed further in Remark \ref{rem:subsegmentsizesregarding}, following the complete description of the phases of the rebalancing algorithm.

\subsection{Examples}
\label{sub:example}
We now provide two examples illustrating our rebalancing algorithm, one corresponding to each of the two transmissions schemes.

\textbf{Example illustrating Scheme 1:}
Consider a database with $K=8$ nodes satisfying the $r$-balanced cyclic storage condition with replication factor $r=6$. A file $W$ is thus split into segments $W_1, \dots, W_8$ such that the segment $W_1$ is stored in nodes $\{1 \boxplus_8 \langle 6\rangle \}$, $W_2$ in nodes $\{2 \boxplus_8 \langle 6\rangle\}$, $W_3$ in nodes $\{3 \boxplus_8 \langle 6\rangle\}$, and so on.

Node $8$ is removed from the system and its contents, namely $W_3$, $W_4$, $W_5$, $W_6$, $W_7$, $W_8$, must be restored. The rebalancing algorithm performs the following steps.

\textit{Splitting}: The splitting is guided by Principles 1 and 2. Each node splits the segments it contains into subsegments as follows: 
    \begin{itemize}
        \item $W_3$ is a corner segment with respect to the removed node $8$. Thus, it is split into two subsegments. The larger is labelled $W_3^{\{1\}}$ and is of size $\frac{12T}{14}$. This is to be merged into $\fW_{3}$ since $|S_3 \cap \tilde{S}_3| = 5 = (r-1)$. The other segment is labelled $W_3^{\{2\}}$ and is of size $\frac{2T}{14}$. As before, the idea is to merge this into $\fW_2$ to maintain a balanced database. 
        \item The other corner segment $W_8$ is handled similarly. It is split into two subsegments labelled $W_8^{\{7\}}$ and $W_8^{\{6\}}$ of sizes $\frac{12T}{14}$ and $\frac{2T}{14}$ respectively.
        
    \item $W_4$ is a middle segment for node $8$. It is split into two subsegments labelled $W_4^{\{2\}}$ and $W_4^{\{3\}}$ of sizes $\frac{10T}{14}$ and $\frac{4T}{14}$ respectively. The intent once again is to merge $W_4^{\{2\}}$ into $\fW_4$ and $W_4^{\{3\}}$ into $\fW_3$ since both $|S_4 \cap \tilde{S}_3| = |S_4 \cap \tilde{S}_4| = 5 = (r-1)$. The remaining middle segments are treated similarly.
        \item $W_5$ into two subsegments labelled $W_5^{\{3\}}$ and $W_5^{\{4\}}$ of sizes $\frac{8T}{14}$ and $\frac{6T}{14}$ respectively.
        \item $W_6$ into two subsegments labelled $W_6^{\{4\}}$ and $W_6^{\{5\}}$ of sizes $\frac{6T}{14}$ and $\frac{8T}{14}$ respectively.
        \item $W_7$ into two subsegments labelled $W_7^{\{5\}}$ and $W_7^{\{6\}}$ of sizes $\frac{4T}{14}$ and $\frac{10T}{14}$ respectively.
        
    \end{itemize}
    The superscript represents the set of nodes to which the subsegment is to be delivered. 
    
\begin{figure}
     \centering
    \includegraphics[scale=0.34]{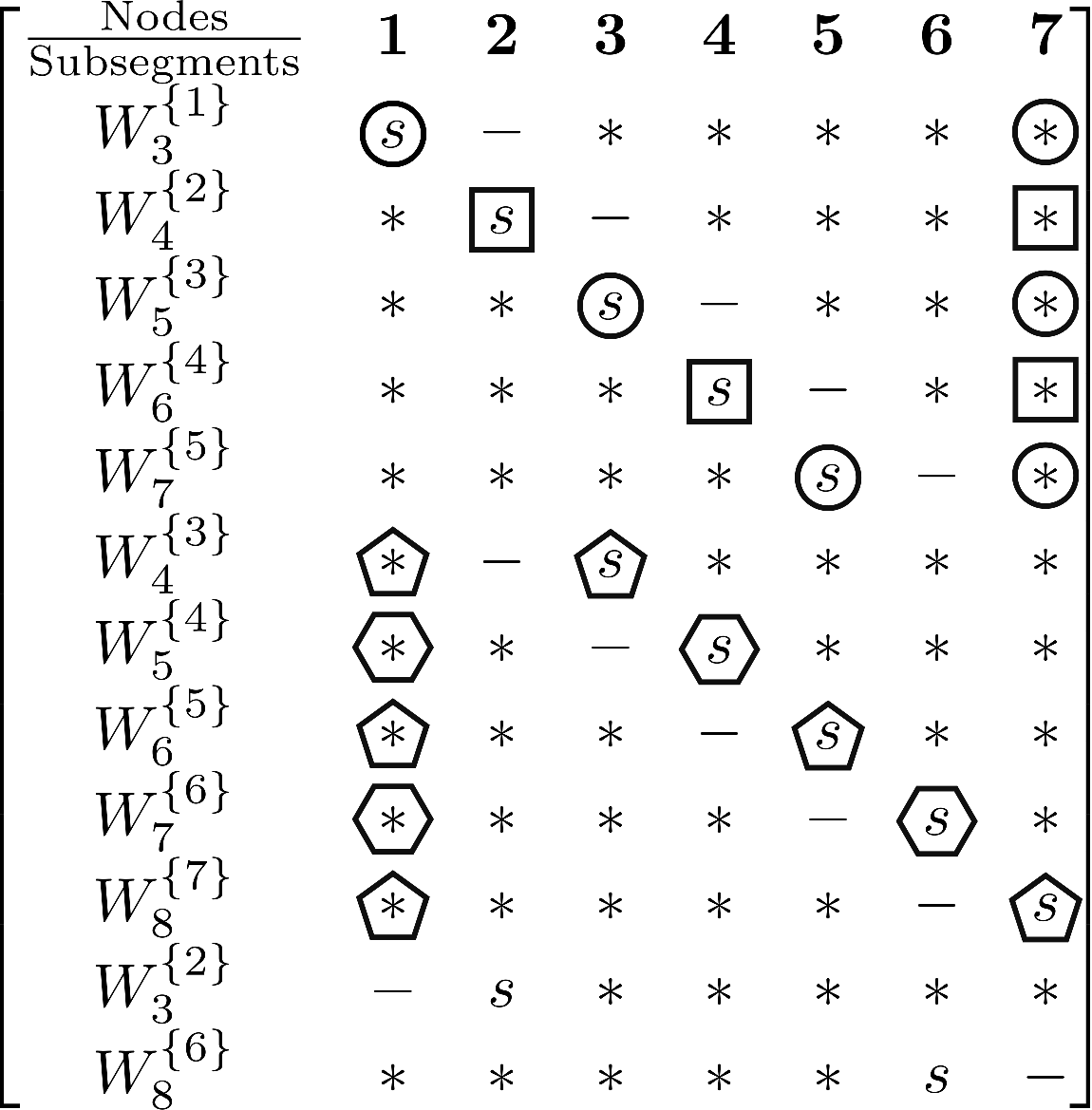}
    \caption{The matrix $M$ for $K=8$, $r = 6$. The rows correspond to subsegments and the columns correspond to nodes. Entry $M_{i,j}$ = `$*$' if the \ith subsegment is contained in the \jth node. $M_{i,j}$ = `$s$' if the \ith subsegment must be delivered to the \jth node. For each shape enclosing an entry, the row and column corresponding each entry with that shape gives a valid XOR-coded transmission.
    }
    \label{fig:pdascheme1}
\end{figure}

\textit{Coding and Transmission}: 
Now, to deliver these subsegments, nodes make use of coded broadcasts. The design of these broadcasts are guided by Principle 3. We elucidate the existence of the `nice' subsets given in Principle 3 using a matrix form (referred to as matrix $M$) in Figure \ref{fig:pdascheme1}. 
We note that this representation is similar to the combinatorial structure defined for coded caching in \cite{yan2017placement}, called a placement delivery array. Consider a submatrix of $M$ described by distinct rows $i_1, \dots, i_l$ and distinct columns $j_1, \dots, j_{l+1}$. If this submatrix is equivalent to the $l\times (l+1)$ matrix 
    \begin{align}
    \label{eqn:submatrix}
    \begin{bmatrix}
    s & * & \dots & * & * \\
    * & s & \dots & * & *\\
    \vdots & \vdots & \ddots & * & *\\
    * & * & \dots & s & * \\
    \end{bmatrix}
    \end{align}
    up to some row/column permutation, then each of the nodes $j_1, \dots, j_l$ can decode their required subsegment from the XOR of the $i_1^{\text{th}}, \dots, i_l^{\text{th}}$ subsegments which can be broadcasted by the $(l+1)^{\text{th}}$ node. To denote the submatrices we make use of shapes enclosing each requirement (represented using an `$s$') in the matrix. For each shape, the row and column corresponding to each `$s$' result in a XOR-coded transmission. Before such XOR-coding, padding the `shorter' subsegments with $0$s to match the length of the longest subsegment would be required. There are other $s$ entries in the matrix $M$ which are not covered by matrices of type (\ref{eqn:submatrix}). These will correspond to uncoded broadcasts.  Thus, we get the following transmissions from the matrix

    \begin{itemize}
        \item Node $1$ pads $W_6^{\{5\}}$ and $W_4^{\{3\}}$ to size $\frac{12T}{14}$ and broadcasts $W_8^{\{7\}} \oplus W_6^{\{5\}} \oplus W_4^{\{3\}}$.
        \item Similarly, Node $7$ pads $W_5^{\{3\}}$ and $W_7^{\{5\}}$ and broadcasts $W_3^{\{1\}} \oplus W_5^{\{3\}} \oplus W_7^{\{5\}}$.
        
        \item Node $1$ pads $W_5^{\{4\}}$ to size $\frac{10T}{14}$ and broadcasts $W_5^{\{4\}} \oplus W_7^{\{6\}}$.
        \item Similarly, Node $7$ pads $W_6^{\{4\}}$ and broadcasts $W_4^{\{2\}} \oplus W_6^{\{4\}}$.
        
        \item Finally, Node $1$ broadcasts $W_8^{\{6\}}$ and Node $7$ broadcasts $W_3^{\{2\}}$. 
    \end{itemize}
    The total communication load incurred in performing these broadcasts is $\frac{1}{T}\left(2. \frac{12T}{14} + 2. \frac{10T}{14} + 2. \frac{2T}{14}\right) = \frac{24}{7}$. Note that the uncoded load for node removal is $6$.
    
\textit{Decoding}: The uncoded subsegments are directly received by the respective nodes. The nodes present in the superscript of the XORed subsegment proceed to decode their respective required subsegment as follows.
    \begin{itemize}
        \item From the transmission $W_8^{\{7\}} \oplus W_6^{\{5\}} \oplus W_4^{\{3\}}$, Node $7$ contains $W_6, W_4$ and can hence recover $W_8^{\{7\}}$ by XORing away the other subsegments. Similarly, Nodes $3$ and $5$ can recover $W_4^{\{3\}}$ and $W_6^{\{5\}}$ respectively. 
        
        \item From the transmission $W_3^{\{1\}} \oplus W_5^{\{3\}} \oplus W_7^{\{5\}}$, Node $1$ contains $W_5, W_7$ and can recover $W_3^{\{1\}}$. Similarly, Nodes $3$ and $5$ can recover $W_5^{\{3\}}$ and $W_7^{\{5\}}$ respectively. 
        
        \item From the broadcast $W_5^{\{4\}} \oplus W_7^{\{6\}}$, Node $4$ contains $W_7$ and can hence recover $W_5^{\{4\}}$. Similarly, Node $6$ can recover $W_7^{\{6\}}$.
        \item From the broadcast $W_4^{\{2\}} \oplus W_6^{\{4\}}$. Node $2$ contains $W_6$ and can hence recover $W_4^{\{2\}}$. Similarly, Node $4$ can recover $W_6^{\{4\}}$.
    \end{itemize}
\textit{Merging and Relabelling}: To restore the cyclic storage condition, each node $k\in[K-1]$ merges and relabels all segments that must be stored in it in the final database. These newly formed subsegments are $\fW_1, \dots, \fW_7$.
    \begin{itemize}
        \item $\fW_1 = W_1 | W_8^{\{6\}}$ of size $1 + \frac{2T}{14} = \frac{8T}{7}$ is obtained at nodes $\{1,2,3,4,5,6\}$.
        \item $\fW_2 = W_2 | W_3^{\{2\}}$ of size $1 + \frac{2T}{14} = \frac{8T}{7}$ is obtained at nodes $\{2,3,4,5,6,7\}$.
        \item $\fW_3 = W_3^{\{1\}} | W_4^{\{3\}}$ of size $\frac{12T}{14} + \frac{4T}{14} = \frac{8T}{7}$ is obtained at nodes $\{3,4,5,6,7,1\}$.
        \item $\fW_4 = W_4^{\{2\}} | W_5^{\{4\}}$ of size $\frac{10T}{14} + \frac{6T}{14} = \frac{8T}{7}$ is obtained at nodes $\{4,5,6,7,1,2\}$.
        \item $\fW_5 = W_5^{\{3\}} | W_6^{\{5\}}$ of size $\frac{8T}{14} + \frac{8T}{14} = \frac{8T}{7}$ is obtained at nodes $\{5,6,7,1,2,3\}$.
        \item $\fW_6 = W_6^{\{4\}} | W_7^{\{6\}}$ of size $\frac{6T}{14} + \frac{10T}{14} = \frac{8T}{7}$ is obtained at nodes $\{6,7,1,2,3,4\}$.
        \item $\fW_7 = W_7^{\{5\}} | W_8^{\{7\}}$ of size $\frac{4T}{14} + \frac{12T}{14} = \frac{8T}{7}$ is obtained at nodes $\{7,1,2,3,4,5\}$.
    \end{itemize}
    After merging and relabelling, each node keeps only the required segments mentioned previously and discards any extra data present. Since each node now stores $6$ segments each of size $\frac{8T}{7}$, the total data stored is still $48T = rNT$. Thus, the cyclic storage condition is satisfied.

\textbf{Example illustrating Scheme 2:}
Consider a database with $K=6$ nodes satisfying the $r$-balanced cyclic storage condition with replication factor $r=3$. A file $W$ is split into segments $W_1, \dots, W_6$ such that the segment $W_1$ is stored in nodes $1,2$ and $3$, $W_2$ in nodes $2,3$ and $4$, $W_3$ in nodes $3,4$ and $5$, $W_4$ in nodes $4,5$ and $6$, $W_5$ in nodes $5,6$ and $1$, and $W_6$ in nodes $6,1$ and $2$.

Suppose the node $6$ is removed from the system. The contents of node $6$, namely $W_4$, $W_5, W_6$ must be restored. To do so, the rebalancing algorithm performs the following steps.

\textit{Splitting}: Again, each node that contains these segments splits them into subsegments as per Principles 1 and 2.
    \begin{itemize}
        \item $W_4$ is a \textit{corner} segment for the removed node $6$. This is split into three subsegments. The largest is labelled $W_4^{\{1\}}$ and is of size $\frac{7T}{10}$. This subsegment is to be merged into $\fW_{4}$ since $|S_4 \cap \tilde{S}_4| = 2 = (r-1)$. Observe that the superscript of $W_4^{\{1\}}$ represents the set of nodes to which the subsegment is to be delivered, i.e., $\fS_4\setminus S_4$. The remaining 2 subsegments are labelled $W_4^{\{3\}}$ and $W_4^{\{2,3\}}$ and are of sizes $\frac{2T}{10}$, and $\frac{T}{10}$ respectively. In order to maintain a balanced database, these are to be merged into $\fW_{3}$ and $\fW_{2}$ respectively.
        \item Similarly, the other \textit{corner} segment $W_6$ is split into three subsegments labelled $W_6^{\{5\}}, W_6^{\{3\}}$ and $W_6^{\{3,4\}}$ of sizes $\frac{7T}{10}, \frac{2T}{10}$, and $\frac{T}{10}$, and are to be merged with $\fW_5$, $\fW_1$, and $\fW_2$ respectively.
        \item $W_5$ is a middle segment of node $6$. It is split into two subsegments labelled $W_5^{\{2\}}$ and $W_5^{\{4\}}$ of size $\frac{5T}{10}$ each. $W_5^{\{2\}}$ is to be merged into $\fW_5$, and $W_5^{\{4\}}$ into $\fW_{4}$, since both $|S_5
        \cap \tilde{S}_5| = |S_5 \cap \tilde{S}_4| = 2 = (r-1)$.
    \end{itemize}
\begin{figure}[htbp]
     \centering
    \includegraphics[scale=0.34]{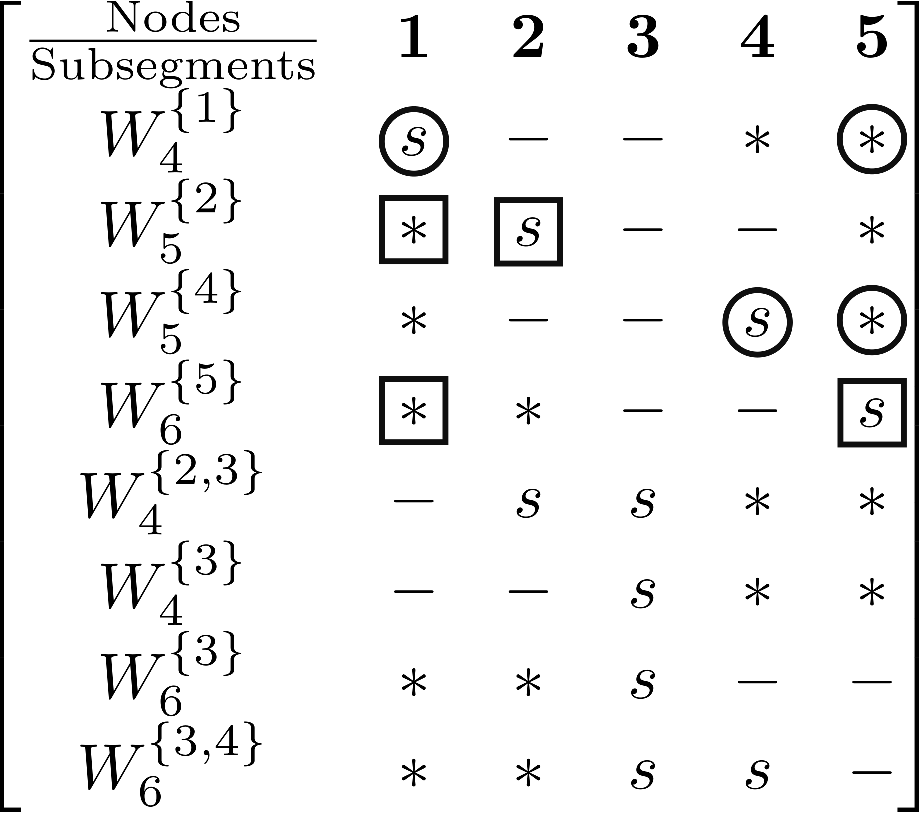}
    \caption{Matrix $M$ for $K=6, r=3$ case. The rows of this matrix $M$ correspond to subsegments and the columns correspond to nodes. Entry $M_{i,j}$ = `$*$' if the \ith  subsegment is contained in the \jth node. $M_{i,j}$ = `$s$' if the \ith subsegment must be delivered to the \jth node. For each shape enclosing an entry, the row and column corresponding each entry with that shape lead to a XOR-coded transmission.}
    \label{fig:pdascheme2}
\end{figure}
\textit{Coding and Transmission}: 
As before, we make use of the placement matrix shown in Fig. \ref{fig:pdascheme2} to explain how nodes perform coded broadcasts as per Principle 3. Consider the submatrix denoted by the circles in Figure \ref{fig:pdascheme2}. This submatrix described by columns $1,4,5$ and rows $1,3$ means that each of the subsegments corresponding to these rows are present in all but one of the nodes corresponding to these columns. Further, node $5$ contains both of these subsegments, and thus node $5$ can broadcast the XOR of them and each of the nodes $1,4$ can recover the respective subsegments denoted by the rows. Further, those $s$ entries in the matrix not covered by any shape lead to uncoded broadcasts. Following these ideas, we get the following transmissions.
    \begin{itemize}
        \item Node $1$ pads $W_5^{\{2\}}$ to size $\frac{7T}{10}$ and broadcasts $W_5^{\{2\}} \oplus W_6^{\{5\}}$.
        \item Similarly, Node $4$ pads $W_5^{\{4\}}$ and broadcasts $W_4^{\{1\}} \oplus W_5^{\{4\}}$.
        \item Finally, Node $1$ broadcasts $W_6^{\{3\}}, W_6^{\{3,4\}}$ and Node $4$ broadcasts $W_4^{\{3\}}, W_4^{\{2,3\}}$.
    \end{itemize}
    The total communication load incurred in performing these broadcasts is 
    $\frac{1}{T} \left( 2.\frac{7T}{10} + 2.\frac{T}{10} + 2.\frac{2T}{10} \right) = 2$. Note that the uncoded rebalancing load for node removal is $3$. 
    
\textit{Decoding}: The uncoded broadcast subsegments are received as-is by the superscript nodes. With respect to any XOR-coded subsegment, the nodes present in the superscript of the subsegment can decode the subsegment, due to the careful design of the broadcasts as per Principle 3. For this example, we have the following.
    \begin{itemize}
        \item From the transmission $W_5^{\{2\}} \oplus W_6^{\{5\}}$, node $2$ can decode $W_5^{\{2\}}$ by XORing away $W_6^{\{5\}}$ and similarly node $5$ can decode $W_6^{\{5\}}$.
        \item Similarly, from $W_4^{\{1\}} \oplus W_5^{\{4\}}$, nodes $1$ and $4$ can recover $W_4^{\{1\}}$ and $W_5^{\{4\}}$ respectively.
    \end{itemize}
\textit{Merging and Relabelling}: To restore the cyclic storage condition,
    each node $k\in[K-1]$ merges and relabels all segments that must be stored in it in the final database. These newly formed subsegments are $\fW_1, \dots, \fW_5$.
    \begin{itemize}
        \item Observe that $W_1$ and $W_6^{\{3\}}$ are available at nodes $\{1,2,3\}$, from either the prior storage or due to decoding. Thus, the segment $\fW_1 = W_1 | W_6^{\{3\}}$ of size $1 + \frac{2T}{10} = \frac{6T}{5}$ is obtained and stored at nodes $\{1,2,3\}$. Similarly, we have the other merge operations as follows. 
        \item $\fW_2 = W_2 | W_4^{\{2,3\}} | W_6^{\{3,4\}}$ of size $1 + \frac{T}{10} + \frac{T}{10} = \frac{6T}{5}$ is obtained at nodes $\{2,3,4\}$.
        \item $\fW_3 = W_3 | W_4^{\{3\}}$ of size $1 + \frac{2T}{10} = \frac{6T}{5}$ is obtained at nodes $\{3,4,5\}$.
        \item $\fW_4 = W_4^{\{1\}} | W_5^{\{4\}}$ of size $\frac{7T}{10} + \frac{5T}{10} = \frac{6T}{5}$ is obtained at nodes $\{4,5,1\}$.
        \item $\fW_5 = W_5^{\{2\}} | W_6^{\{5\}}$ of size $\frac{5T}{10} + \frac{7T}{10} = \frac{6T}{5}$ is obtained at nodes $\{5,1,2\}$.
    \end{itemize}
    After merging and relabelling, each node keeps only the required segments mentioned previously and discards any extra data present. Since each node now stores $3$ segments each of size $\frac{6T}{5}$, the total data stored is still $18T = rNT$. Thus the cyclic storage condition is satisfied.

\subsection{Algorithm for Rebalancing for Node-Removal}
\label{sub:algorithm}
\begin{figure}[H]
\includegraphics[scale=0.3]{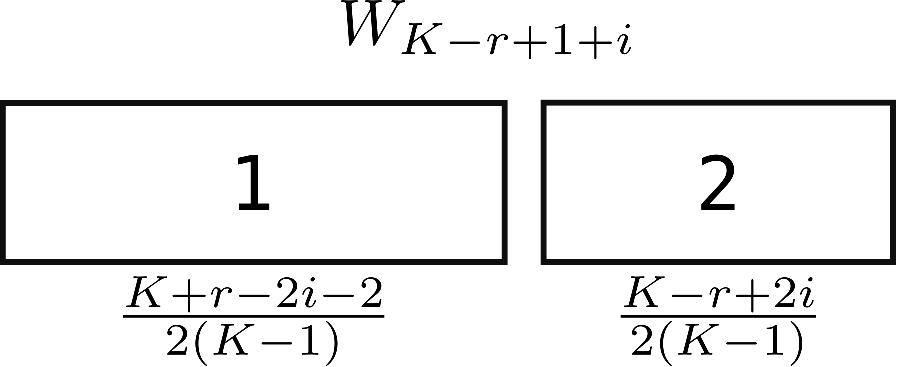}
\centering
\caption{Splitting of middle segments: For each $i \in [r-2]$, $W_{K-r+1+i}$ is split into two parts labelled $W_{K-r+1+i}^{\{i+1\}}$ and $W_{K-r+1+i}^{\{i+K-r\}}$ of sizes $\frac{K+r-2i-2}{2(K-1)}$ and $\frac{K-r+2i}{2(K-1)}$ respectively.}
\label{fig:split1}
\end{figure}

\begin{figure}[H]
\includegraphics[scale=0.3]{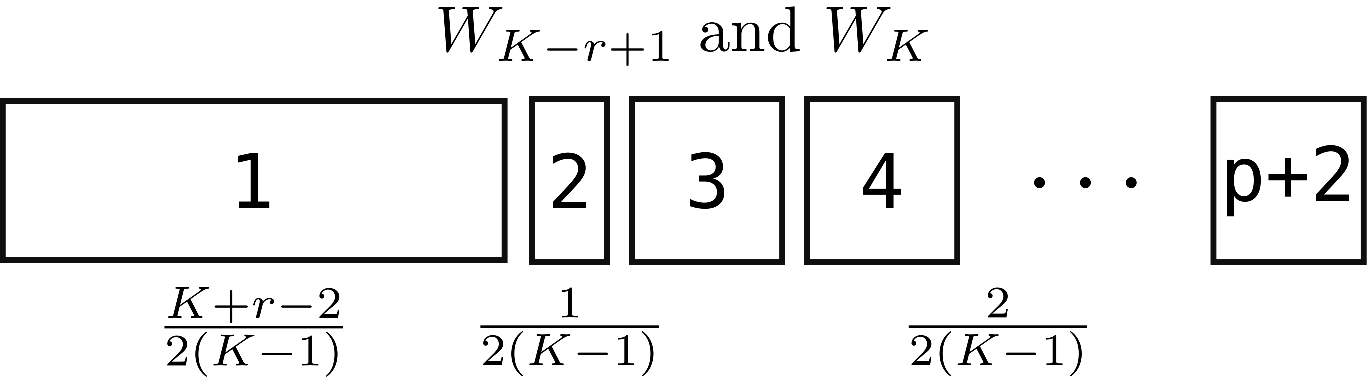}
\centering
\caption{Splitting of corner segments: Let $p=\floor{\frac{K-r}{2}}$. When $K-r$ is odd, $W_{K-r+1}$ is split into $p+2$ parts labelled $W_{K-r+1}^{\{1\}}$, $W_{K-r+1}^{\{(K-r-p) \boxplus_{K-1} \langle \min(r,p+1) \rangle\}}$, and $W_{K-r+1}^{\{(K-r+1-j) \boxplus_{K-1} \langle \min(r,j) \rangle\}}$ for $j = 1, \dots, p$; of sizes $\frac{K+r-2}{2(K-1)}, \frac{1}{2(K-1)}, \frac{2}{2(K-1)}, \frac{2}{2(K-1)}, \dots, \frac{2}{2(K-1)}$. Similarly, $W_{K}$ is split into $p+2$ parts labelled $W_{K}^{\{K-1\}}$, $W_{K}^{\{(r+p) \boxminus_{K-1} \langle \min(r,p+1) \rangle\}}$, and $W_{K}^{\{(r-1+j) \boxminus_{K-1} \langle \min(r,j) \rangle\}}$ for $j = 1, \dots, p$; of sizes $\frac{K+r-2}{2(K-1)}, \frac{1}{2(K-1)}, \frac{2}{2(K-1)}, \frac{2}{2(K-1)}, \dots, \frac{2}{2(K-1)}$ respectively.}
\label{fig:split2}
\end{figure}

\begin{figure}[H]
\includegraphics[scale=0.3]{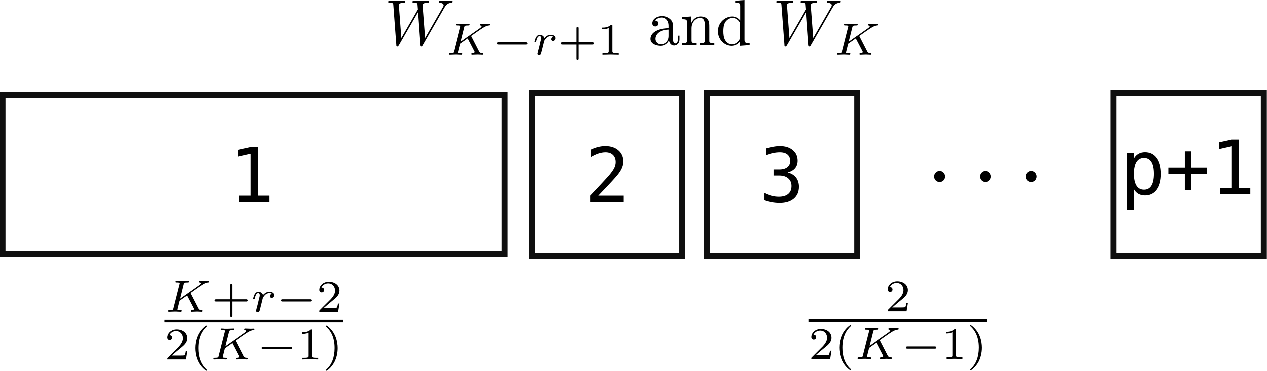}
\centering
\caption{Let $p=\floor{\frac{K-r}{2}}$. When $K-r$ is even, $W_{K-r+1}$ is split into $p+1$ parts labelled $W_{K-r+1}^{\{1\}}$, and $W_{K-r+1}^{\{(K-r+1-j) \boxplus_{K-1} \langle \min(r,j) \rangle\}}$ for $j = 1, \dots, p$; of sizes $\frac{K+r-2}{2(K-1)}, \frac{2}{2(K-1)}, \frac{2}{2(K-1)}, \dots, \frac{2}{2(K-1)}$. Similarly, $W_{K}$ is split into $p+1$ parts labelled $W_{K}^{\{K-1\}}$, and $W_{K}^{\{(r-1+j) \boxminus_{K-1} \langle \min(r,j) \rangle\}}$ for $j = 1, \dots, p$; of sizes $\frac{K+r-2}{2(K-1)}, \frac{2}{2(K-1)}, \frac{2}{2(K-1)}, \dots, \frac{2}{2(K-1)}$ respectively.}
\label{fig:split3}
\end{figure}

In this section, following the intuition built in Subsection \ref{sub:intuition}, we give our complete rebalancing algorithm (Algorithm \ref{algo:rebalancing}) for removal of a node from an $r$-balanced cyclic database on $K$ nodes (with $r\in \{3,\hdots,K-1\}$). We thus prove the node-removal result in Theorem \ref{thm:achieve}. Algorithm \ref{algo:rebalancing} initially invokes the SPLIT routine (described in Algorithm \ref{algo:split}) which gives the procedure to split segments into subsegments. Each subsegment's size is assumed to be an integral multiple of $\frac{1}{2(K-1)}$. This is without loss of generality, by the condition on the size $T$ of each segment as in Theorem \ref{thm:achieve}. This splitting scheme is also illustrated in the figures Fig. \ref{fig:split1}-\ref{fig:split3}. Guided by Claim \ref{claim:threshold}, based on the value of $r$, Algorithm \ref{algo:rebalancing} selects between two routines that correspond to the two transmission schemes: SCHEME 1 and SCHEME 2, both of which are essentially clique-cover-based index coding schemes \cite{ISCOD_BiK}. These schemes are given in Algorithm \ref{algo:case1} and \ref{algo:case2}. We note that, since the sizes of the subsegments may not be the same after splitting, appropriate zero-padding (up to the size of the larger subsegment) is done before the XOR operations are performed in the two schemes. Finally, the MERGE routine (given in Algorithm \ref{algo:merge}) is run at the end of Algorithm \ref{algo:rebalancing}. This merges the subsegments and relabels the merged segments as the target segments, thus resulting in the target $r$-balanced cyclic database on $K-1$ nodes.

\begin{remark}[Complexity of Algorithm \ref{algo:rebalancing}]
\label{remark:compcomplexitynoderemoval}
    The computational complexity of Algorithm \ref{algo:rebalancing} is proved in Appendix \ref{app:computationcomplexitynoderemoval} to be $O(K^3T)$. 
\end{remark}

\begin{remark}[Rebalancing from removal of arbitrary node]
\label{remark:arbitrarynoderemoval}Note that, for ease of understanding, we describe the algorithms for the removal of node $K$ from the system. The scheme for the removal of a general node $i$ can be obtained as follows. Consider the set of permutations $\phi_i: [K] \rightarrow [K]$, where $\phi_i(j) = j \boxminus_K (K - i)$, for $i \in [K]$. If a node $i$ is removed instead of $K$, we replace each label $j$ in the subscript and superscript of the subsegments in our Algorithms \ref{algo:split}-\ref{algo:merge} with $\phi_i(j)$. Due to the cyclic nature of both the input and target databases,  we naturally obtain the rebalancing scheme for the removal of node $i$. 
\end{remark}

\begin{remark}[Intuition regarding the splitting scheme]
\label{rem:subsegmentsizesregarding}
    We comment regarding the intuition behind the size of the subsegments generated by the splitting scheme, as shown in Fig. \ref{fig:split1}, \ref{fig:split2}, and \ref{fig:split3}. These sizes were essentially identified by generalizing a number of numerical examples constructed for a variety of parameter choices, while adhering to the following rationale. 
    \begin{itemize}
        \item Uncoded transmissions are needed for `middle' nodes that are `far-away' from the removed node $K$, specifically some subset of the nodes $\{\min(K-r,r),\hdots,\max(K-r,r)\}$. This is because, these nodes have either less content or  no content, that is common with the content stored in removed node $K$. The subsegments intended for the middle-nodes adjacent to $K-r$ are chosen from some subsegments of the corner segment $W_{K-r+1}$ (because of the \textit{Principle 2}). 
        Similarly, the subsegments for middle-nodes close to $r$ are some subsegments of $W_K$ (again, this is due to \textit{Principle 2}). These uncoded transmissions must intuitively be small-sized, in order to keep the rebalancing load small. The small subsegments shown in Fig. \ref{fig:split2} and \ref{fig:split3} illustrate this aspect. The transmission schemes indicate that these small subsegments are sent as uncoded broadcasts. 
        \item The other subsegments, i.e., the large subsegments of the corner segments, and the pairs of subsegments of the middle segments, participate in coded transmissions. Observe from Fig. \ref{fig:split1}, \ref{fig:split2}, and \ref{fig:split3}, that all these subsegments are meant for the nodes $\{1,2,\hdots,r-1\}\cup\{K-1,K-2,\hdots,K-r+1\}$ only. These nodes are `nearer' to the removed node $K$ and thus share common segments with $K$. This creates coding opportunities, thus enabling the use of coded transmissions in our schemes. Since these can be part of coded transmissions, their sizes can be larger than those involved in uncoded transmissions.
        \item With the above two intuitive reasons, we also take into account the merge process to fix the precise subsegment sizes. Observe that the new segments are created as per Algorithm \ref{algo:merge} by merging old segments (or parts of the old segments) with the new subsegments. Each new segment formed must be of size $K/(K-1)$. From the correctness of the rebalancing algorithm in obtaining the final target database (verified in Subsection \ref{sub:correctness}), we observe that the specific choices of each of the subsegments in the splitting process satisfies this requirement as well.  
    \end{itemize}
\end{remark}

\begin{algorithm}[H]
\caption{Rebalancing Scheme for Node Removal from Cyclic Database}
\label{algo:rebalancing}
\begin{algorithmic}[1]
\Procedure{Transmit}{}
    \State \textsc{Split()} \Comment{Call \textsc{Split}}
     \If{$r \geq r_{th}=\ceil{\frac{2K+2}{3}}$}
        \State \textsc{Scheme 1()} \Comment{Call \textsc{Scheme 1}}
    \Else
        \State \textsc{Scheme 2()} \Comment{Call \textsc{Scheme 2}}
    \EndIf
    \State \textsc{Merge()} \Comment{Call \textsc{Merge}}
\EndProcedure
\end{algorithmic}
\end{algorithm}

\begin{algorithm}[H]
\caption{Splitting Scheme}
\label{algo:split}
\begin{algorithmic}[1]
\Procedure{Split}{}
\For{each $i \in [r-2]$}
        \State Split $W_{K-r+1+i}$ into subsegments labelled $W_{K-r+1+i}^B$ for $B \in \{\{i+1\}, \{i+K-r\} \}$, where the size of the subsegment is $\begin{cases}
        \frac{K+r-2i-2}{2(K-1)}, & \text{ if } B = \{i+1\} \\
        \frac{K-r+2i}{2(K-1)}, & \text{ if } B = \{i+K-r\}
        \end{cases}
        $
    \EndFor
    \If{$K-r$ is odd}
        \State Let  $p = \floor{\frac{K-r}{2}}$
        \State Split $W_{K-r+1}$ into $p+2$ subsegments labeled $W_{K-r+1}^B$, for $B \in \{\{1\},\{(K-r-p) \boxplus_{K-1} \langle \min(r,p+1) \rangle\} \} \cup \{\{(K-r+1-j) \boxplus_{K-1} \langle \min(r,j) \rangle\} : j \in [p] \}$
        where the size of the subsegment is $\begin{cases}
        \frac{K+r-2}{2(K-1)}, & \text{ if } B = \{1\} \\
        \frac{1}{2(K-1)}, & \text{ if } B = \{(K-r-p) \boxplus_{K-1} \langle \min(r,p+1) \rangle\} \\
        \frac{2}{2(K-1)}, & \text{ otherwise}
        \end{cases}
        $
        \State Split $W_{K}$ into $p+2$ subsegments labelled $W_{K}^B$, for $B \in \{\{K-1\}, \{(r+p) \boxminus_{K-1} \langle \min(r,p+1) \rangle\}\} \cup \{\{(r-1+j) \boxminus_{K-1} \langle \min(r,j) \rangle\} : j \in [p]\}$ where the size of the subsegment is $\begin{cases}
        \frac{K+r-2}{2(K-1)}, & \text{ if } B = \{K-1\} \\
        \frac{1}{2(K-1)}, & \text{ if } B = \{(r+p) \boxminus_{K-1} \langle \min(r,p+1) \rangle\} \\
        \frac{2}{2(K-1)}, & \text{ otherwise}
        \end{cases}
        $
    \Else 
        \State Let  $p = \frac{K-r}{2}$
        \State Split $W_{K-r+1}$ into $p+1$ subsegments labelled $W_{K-r+1}^B$, for $B \in \{1\} \cup \{\{(K-r+1-j) \boxplus_{K-1} \langle \min(r,j) \rangle\} : j \in [p]\}$ where the size of the subsegment is $\begin{cases}
        \frac{K+r-2}{2(K-1)}, & \text{ if } B = \{1\} \\
        \frac{2}{2(K-1)}, & \text{ otherwise}
        \end{cases}
        $
        \State Split $W_{K}$ into $p+1$ subsegments labelled $W_{K}^B$, for $B \in \{K-1\} \cup \{\{(r-1+j) \boxminus_{K-1} \langle \min(r,j) \rangle\} : j \in [p]\}$ where the size of the subsegment is $\begin{cases}
        \frac{K+r-2}{2(K-1)}, & \text{ if } B = \{K-1\} \\
        \frac{2}{2(K-1)}, & \text{ otherwise}
        \end{cases}
        $
    \EndIf
\EndProcedure
\end{algorithmic}
\end{algorithm}

\begin{algorithm}[H]
\caption{Transmission Scheme 1}
\label{algo:case1}
\begin{algorithmic}[1]
\Procedure{Scheme 1}{}
    \For{each $i= 1, \dots, K-r$}
        \State Node $1$ broadcasts $\bigoplus_{j=0}^{\floor{\frac{r-1-i}{K-r}}} W_{K+1-i-j(K-r)}^{\{K-i-j(K-r)\}}$.
        \State Node $K-1$ broadcasts $\bigoplus_{j=0}^{\floor{\frac{r-1-i}{K-r}}} W_{K-r+i+j(K-r)}^{\{i+j(K-r)\}}$
    \EndFor
    \State Node $1$ broadcasts all subsegments of $W_K$ except $W_K^{\{K-1\}}$.
    \State Node $K-1$ broadcasts all subsegments of $W_{K-r+1}$ except  $W_{K-r+1}^{\{1\}}$.  
\EndProcedure
\end{algorithmic}
\end{algorithm}

\begin{algorithm}[H]
\caption{Transmission Scheme 2}
\label{algo:case2}
\begin{algorithmic}[1]

\Procedure{Scheme 2}{}
    \For{each $i = 2, \dots, r-1$}
        \State Node $1$ broadcasts $W_{K-r+i}^{\{i\}} \oplus W_{K-r+i+1}^{\{K-r+i\}}$. 
    \EndFor
    \State Node $K-1$ broadcasts $W_{K-r+1}^{\{1\}} \oplus W_{K-r+2}^{\{K-r+1\}}$. 
    \State Node $1$ broadcasts all subsegments of $W_K$ except $W_K^{\{K-1\}}$.
    \State Node $K-1$ broadcasts all subsegments of $W_{K-r+1}$ except $W_{K-r+1}^{\{1\}}$.  
\EndProcedure
\end{algorithmic}
\end{algorithm}

\begin{algorithm}[H]
\caption{Merging and Relabelling}
\label{algo:merge}
\begin{algorithmic}[1]
\Procedure{Merge}{}
    \For{each $i= 1, \dots, r-1$}
        \State Each node in $\{(K-r+i) \boxplus_{K-1} \langle r \rangle\}$ performs the concatenation $\fW_{K-r+i} = W_{K-r+i}^{\{i\}} | W_{K-r+i+1}^{\{K-r+i\}}$.
    \EndFor
    \If{$K-r$ is even}
        \For{each $i= 1, \dots, \frac{K-r}{2}$}
            \State Each node in $\{i \boxplus_{K-1} \langle r \rangle\}$ performs the concatenation $\fW_{i} = W_{i} | W_{K}^{\{(r-1+i) \boxminus_{K-1} \langle min(r,i) \rangle\}}$.
        \EndFor
        \For{each $i= \frac{K-r}{2}+1, \dots, K-r$}
            \State Each node in $\{i \boxplus_{K-1} \langle r \rangle\}$ performs the concatenation $\fW_{i} = W_{i} | W_{K-r+1}^{\{i \boxplus_{K-1} \langle min(r,K-r-i+1) \rangle\}}$.
        \EndFor
    \Else
        \For{each $i= 1, \dots, \frac{K-r-1}{2}$}
            \State Each node in $\{i \boxplus_{K-1} \langle r \rangle\}$ performs the concatenation $\fW_{i} = W_{i} | W_{K}^{\{(r-1+i) \boxminus_{K-1} \langle min(r,i) \rangle\}}$.
        \EndFor
        \For{each $i= \frac{K-r+1}{2}+1, \dots, K-r$}
            \State Each node in $\{i \boxplus_{K-1} \langle r \rangle\}$ performs the concatenation $\fW_{i} = W_{i} | W_{K-r+1}^{\{i  \boxplus_{K-1} \langle min(r,K-r-i+1) \rangle\}}$.
        \EndFor
        \State Each node in $\left\{ \left(\frac{K-r+1}{2} \right) \boxplus_{K-1} \langle r \rangle \right \}$ performs the concatenation $\fW_{\frac{K-r+1}{2}} = W_{\frac{K-r+1}{2}} |  \newline W_{K}^{\{(r+p)  \boxminus_{K-1} \langle \min(r,p+1) \rangle\}} | W_{K-r+1}^{\{(K-r-p)  \boxplus_{K-1} \langle \min(r,p+1) \rangle\}}$, where $p = \floor{\frac{K-r}{2}}$.
    \EndIf
\EndProcedure
\end{algorithmic}
\end{algorithm}

\textbf{Note:} Once the target segments $\fW_1, \dots, \fW_{K-1}$ are recovered at the required nodes, any extra bits present at the node are discarded.

\subsection{Correctness of Algorithm \ref{algo:rebalancing}}
\label{sub:correctness}
To check the correctness of the scheme, we have to check the correctness of the encoding, decoding, and the merging. It is straightforward to check that the nodes that broadcast any transmission, whether coded or uncoded subsegments, contain all respective subsegments according to the design of the initial storage. Thus, the XOR-coding and broadcasts given in the transmissions schemes are correct. For checking the decoding, we must check that each subsegment can be decoded at the corresponding `superscript' nodes  where it is meant to be delivered. We must also check that the merging scheme is successful, i.e., at any node, all the subsegments to be merged into a target segment are available at that node. Finally, we check that the target database is the cyclic database on $K-1$ nodes.

Now, we focus on checking the decoding of the transmissions in both Scheme 1 and Scheme 2. Clearly, all uncoded transmissions are directly received. Thus, we now check only the decoding involved for XOR-coded transmissions, for the two schemes. 
\begin{itemize}
    \item \textbf{Decoding for Scheme 1: }For each $i \in [K-r]$, two broadcasts $\bigoplus_{j=0}^{\floor{\frac{r-1-i}{K-r}}} W_{K+1-i-j(K-r)}^{\{K-i-j(K-r)\}}$ and $\bigoplus_{j=0}^{\floor{\frac{r-1-i}{K-r}}} W_{K-r+i+j(K-r)}^{\{i+j(K-r)\}}$ are made. 
    Consider the first broadcast. Let $J = \{0, \dots, \floor{\frac{r-1-i}{K-r}}\}$. For some $j \in J$, consider the segment $W_{K+1-i-j(K-r)}$. For any $j' \in J \backslash \{j\}$, we claim that node $K-i-j'(K-r)$ contains the segment $W_{K+1-i-j(K-r)}$. Going through all possible $j,j'$ would then mean that all the segments in this first XOR-coded broadcast can be decoded at the respective superscript-nodes. Now, for node $K-i-j'(K-r)$ to contain the segment $W_{K+1-i-j(K-r)}$, the following condition must be satisfied: 
    \begin{itemize} 
    \item Condition (A): $K-i-j'(K-r) \in S_{K+1-i-j(K-r)}=\{(K+1-i-j(K-r)) \boxplus_K \langle r \rangle\}$.
    \end{itemize} To remove the wrap-around, we simplify Condition (A) into two cases based on the relation between $j$ and $j'$. For Condition (A) to hold, it is easy to check that one of the following pairs of inequalities must hold.
    \begin{enumerate}
        \item if $j < j'$, $K+1-i-j(K-r) \leq 2K-i-j'(K-r) \leq K+1-i-j(K-r)+r-1$
         \item  if $j > j'$, $ K+1-i-j(K-r) \leq K-i-j'(K-r) \leq K+1-i-j(K-r)+r-1$.
    \end{enumerate}
    
    Consider the first inequality. First we prove that when $j < j'$, $K+1-i-j(K-r) \leq 2K-i-j'(K-r)$. To show this, we consider the following sequence of equations,
    \begin{align*}
    (2K&-i-j'(K-r))
    -(K+1-i-j(K-r)) \\
    &= K-1 - (j'-j)(K-r) \\ 
    &\stackrel{(a)}{\geq}
    K-1 -
    \left(
    \frac{r-1-i}{K-r}
    \right)(K-r) \\
    &\geq K-r+i \\
    &\stackrel{(b)}{\geq} K-r+1 \\
    &\stackrel{(c)}{\geq} K-(K-1)+1 \\
    &\geq 0.
    \end{align*}
    where (a) holds as the maximum value of $j'-j$ is equal to $\floor{\frac{r-1-i}{K-r}}$, (b) holds as the minimum value of $i$ is 1, and (c) holds as the maximum value of $r$ is $K-1$. 
    
    Similarly,
    \begin{align*}
    &(K+1-i-j(K-r)+r-1) \\
    &\quad - (2K-i-j'(K-r)) \\
    &= r-K + (j'-j)(K-r) \\ 
    &\stackrel{(a)}{\geq}
    r-K + (K-r) \\
    &= 0.
    \end{align*}
    where (a) holds as the minimum value of $j'-j$ is equal to 1. 
    
    Now, for the second inequality, we first prove that when $j > j'$, $ K+1-i-j(K-r) \leq K-i-j'(K-r)$. To show this, we consider the following sequence of equations
    \begin{align*}
    &(K-i-j'(K-r)) \\
    &\quad -(K+1-i-j(K-r)) \\
    &= (j-j')(K-r)-1 \\ 
    &\stackrel{(a)}{\geq} K-r-1 \\
    &\stackrel{(b)}{\geq} K-(K-1)-1 \\
    &= 0.
    \end{align*}
    where (a) holds as the minimum value of $j-j'$ is equal to 1 and (b) holds as the maximum value of $r$ is $K-1$. Similarly,
    \begin{align*}
    &(K+1-i-j(K-r)+r-1) \\
    &\quad - (K-i-j'(K-r)) \\
    &= r -(j-j')(K-r) \\ 
    &\stackrel{(a)}{\geq}
    r -
    \left(
    \frac{r-1-i}{K-r}
    \right)(K-r) \\
    &\geq 1+i \\
    &\stackrel{(b)}{\geq} 0.
    \end{align*}
    where (a) holds as the maximum value of $j-j'$ is equal to $\floor{\frac{r-1-i}{K-r}}$ and (b) holds as the minimum value of $i$ is 1.
    
    Hence, all the inequalities for both the cases are true. Similar arguments hold for the second broadcast as well.

    \item \textbf{Decoding for Scheme 2: } For each $i \in [r-1]$, a broadcast $W_{K-r+i}^{\{i\}} \oplus W_{K-r+i+1}^{\{K-r+i\}}$ is made (c.f. Lines 3,5 in Algorithm \ref{algo:case2}). Now, node $i$ contains the subsegment $W_{K-r+i+1}$ since $(K-r+i+1) \boxplus_K (r-1) = i$. Similarly, node $K-r+i$ clearly contains the subsegment $W_{K-r+i}$. Thus, node $i$ can decode $W_{K-r+i}^{\{i\}}$ and node $K-r+i$ can decode $W_{K-r+i+1}^{\{K-r+i\}}$. Thus, we have verified the correctness of the transmission schemes. 
    \item \textbf{Checking the merging phase: }  Initially, for each $i \in [K]$, $W_i$ is stored at the nodes $\{ i \boxplus_K \langle r \rangle \}$. After the transmissions are done, $\fW_{i}$ is obtained by merging some subsegments with $W_i$, for each $i \in \{1, \dots, K\}$. Now, to verify that the merging can be done correctly, we need to show that all these subsegments are present at the nodes  $\{ i \boxplus_{K-1} \langle r \rangle \}$ after the transmissions are done.
    \begin{itemize}
        \item For each $i \in [r-1]$, consider the segment $\fW_{K-r+i}$ which is obtained by merging $W_{K-r+i}^{\{i\}}$ with $W_{K-r+i+1}^{\{K-r+i\}}$, (c.f. Algorithm \ref{algo:merge} Line 2 and 3). We observe that $W_{K-r+i}^{\{i\}}$ was present in $\{(K-r+i) \boxplus_{K-1} \langle r \rangle \} \backslash \{i\}$ before rebalancing and was decoded by node $i$ during the rebalancing process. Similarly,  $W_{K-r+i+1}^{\{K-r+i\}}$ was present in $\{(K-r+i) \boxplus_{K-1} \langle r \rangle \} \backslash \{K-r+i\}$ before rebalancing and was decoded by node $K-r+i$ during the rebalancing process.

        \item For each $i \in \{1,\dots,\floor{\frac{K-r}{2}}\}$, $\fW_{i}$ is obtained by merging $W_{i}$ and $W_{K}^{\{(r-1+i) \boxminus_{K-1} \langle \min(r,i) \rangle\}}$, (c.f. Algorithm \ref{algo:merge} Lines 7 and 14). Each node in $\{i \boxplus_{K-1} \langle r \rangle\}$ that does not contain $W_K$ can obtain $W_{K}^{\{(r-1+i) \boxminus_{K-1} \langle \min(r,i) \rangle\}}$ using the broadcasts made in Lines 6-7 of Algorithms \ref{algo:case1} and \ref{algo:case2}. Thus, $\fW_i$ can be obtained at the nodes $\{ i \boxplus_{K-1} \langle r \rangle \}$. 
        
        \item We use similar arguments for the target segments $\fW_i$ for each $i \in \{\ceil{\frac{K-r}{2}}+1, \dots, K-r\}$ and $\fW_{\frac{K-r+1}{2}}$, if $K-r$ is odd.
    \end{itemize}
    
    \item \textbf{Checking the target database structure:} As we have already checked the placement of the new subsegments in the previous bullet, we only have to show that the sizes of all the segments after rebalancing are equal. For this, we look at how the new segments are formed by merging some subsegments with the older segment. 
    \begin{itemize}
        \item The size of the target segment $\fW_{K-r+i}$, for each $i \in \{1, \dots, r-1\}$, is $\frac{K+r-2(i-1)-2}{2(K-1)} + \frac{K-r+2i}{2(K-1)} = \frac{K}{K-1}$ (c.f. Algorithm \ref{algo:merge} - Line 2 and 3).
        \item The size of the target segment $\fW_{i}$, for each $i \in [\floor{\frac{K-r}{2}}]$, is $1 + \frac{2}{2(K-1)} = \frac{K}{K-1}$ (c.f. Algorithm \ref{algo:merge} - Line 6, 7, 13, and 14).
        \item For even $K-r$, the size of the target segment $\fW_{i}$, for each $i \in \{\frac{K-r}{2}+1, \dots, K-r\}$, is $1 + \frac{2}{2(K-1)} = \frac{K}{K-1}$ (c.f. Algorithm \ref{algo:merge} - Line 9 and 10).
        \item For odd $K-r$, the size of the target segment $\fW_{i}$, for each $i \in \{\frac{K-r+1}{2}+1, \dots, K-r\}$, is $1 + \frac{2}{2(K-1)} = \frac{K}{K-1}$ (c.f. Algorithm \ref{algo:merge} - Line 16 and 17).
        \item The size of the target segment $\fW_{\frac{K-r+1}{2}}$ is $1 + \frac{1}{2(K-1)} + \frac{1}{2(K-1)} = \frac{K}{K-1}$. (c.f. Algorithm \ref{algo:merge} - Line 19).
    \end{itemize}
    We can see that the sizes of all the segments after rebalancing are the same, i.e., $\frac{K}{K-1}$. This completes the verification of the correctness of our rebalancing algorithm, which assures that the target database is an $r$-balanced cyclic database on $K-1$ nodes.
    
\end{itemize}


\subsection{Communication Load of Algorithm \ref{algo:rebalancing}}
\label{subsec:comm_load_noderemoval}
We now calculate the communication loads of the two schemes. 
For the uncoded broadcasts in both schemes 
corresponding to lines 6, 7 in both Algorithms \ref{algo:case1} and \ref{algo:case2}, the communication load incurred is $2 \left( \frac{K-r-1}{2}.\frac{2}{2(K-1)} + \frac{1}{2(K-1)}\right) = \frac{K-r}{(K-1)}$, when $K-r$ is odd, and $2 \left( \frac{K-r}{2}.\frac{2}{2(K-1)}\right) = \frac{K-r}{(K-1)}$ when $K-r$ is even, respectively. Now, we analyse the remainder of the communication loads of the two schemes. 

\subsubsection{Scheme 1}
The coded broadcasts made in Scheme 1 are $\bigoplus_{j=0}^{\floor{\frac{r-1-i}{K-r}}} W_{K+1-i-j(K-r)}^{\{K-i-j(K-r)\}}$ and $\bigoplus_{j=0}^{\floor{\frac{r-1-i}{K-r}}} W_{K-r+i+j(K-r)}^{\{i+j(K-r)\}}$ for each $i \in [K-r]$ (c.f. Algorithm \ref{algo:case1} Lines 2-5). Once again, the subsegments involved in the broadcast are padded so that they are of the same size. Consider $\bigoplus_{j=0}^{\floor{\frac{r-1-i}{K-r}}} W_{K-r+i+j(K-r)}^{\{i+j(K-r)\}}$. The size of each subsegment involved is $\frac{K+r-2(i+j(K-r)-1)-2}{2(K-1)}$. Thus, the subsegment having the maximum size is the one corresponding to $j=0$ having size $\frac{K+r-2i}{2(K-1)}$. Similarly, each subsegment involved in $\bigoplus_{j=0}^{\floor{\frac{r-1-i}{K-r}}} W_{K+1-i-j(K-r)}^{\{K-i-j(K-r)\}}$ is of size $\frac{K-r+2(r-i-j(K-r))}{2(K-1)}$. Thus, the maximum size is the one corresponding to $j=0$ having size, $\frac{K+r-2i}{2(K-1)}$. Therefore, the communication load is 
\begin{align*}
   L_1(r) &= 2 \cdot \sum_{i=1}^{K-r} \frac{K+r-2i}{2(K-1)} \\
    &= \left(\frac{1}{K-1}\right) \left((K-r)(K+r) - (K-r)(K-r+1)\right) \\
    &= \frac{(K-r)(2r-1)}{(K-1)}.
\end{align*}

\subsubsection{Scheme 2}
The coded broadcasts made in Scheme 2 involve $W_{K-r+i}^{\{i\}} \oplus W_{K-r+i+1}^{\{K-r+i\}}$ for each $i \in [r-1]$ (c.f. Algorithm \ref{algo:case2} Lines 2-5). Since the smaller subsegment of each pair is padded to match the size of the larger, the cost involved in making the broadcast depends on the larger subsegment. Now, the size of these subsegments are $\frac{K+r-2j-2}{2(K-1)}$ and $\frac{K-r+2(j+1)}{2(K-1)}$ respectively (c.f. Figures \ref{fig:split1}, \ref{fig:split2}, \ref{fig:split3}). For the first subsegment to be larger, $\frac{K+r-2j-2}{2(K-1)} \geq \frac{K-r+2(j+1)}{2(K-1)}$. It is easy to verify that this occurs when $j \leq \frac{r}{2}-1$. We separate our analysis into cases based on the parity of $r$.

For odd $r$, we have, 
\begin{align*}
    L_2(r)
    &= \sum_{j=0}^{\frac{r-1}{2}-1}
    \frac{K+r-2j-2}{2(K-1)} +
    \sum_{j=\frac{r-1}{2}}^{r-2}
    \frac{K-r+2(j+1)}{2(K-1)} \\
    &\stackrel{(a)}{=}
    \sum_{j=0}^{\frac{r-1}{2}-1}
    \frac{K+r-2j-2}{2(K-1)} \\
    &\quad +
    \sum_{j'=0}^{\frac{r-1}{2}-1}
    \frac{K-r+2(r-2-j'+1)}{2(K-1)} \\
    &=
    2 \sum_{j=0}^{\frac{r-1}{2}-1}
    \frac{K+r-2j-2}{2(K-1)} \\
    &=
    \left(
    \frac{2}{2(K-1)}
    \right)
    \Bigg(
    \left(\frac{r-1}{2}\right)(K+r-2) \\
    &\qquad -
    \left(\frac{r-1}{2}-1\right)
    \left(\frac{r-1}{2}\right)
    \Bigg) \\
    &=
    \left(
    \frac{1}{2(K-1)}
    \right)
    (r-1)
    \left(
    K+\frac{r-1}{2}
    \right),
\end{align*}
where $(a)$ is obtained by changing the variable $j' = (r-2)-j$.

Similarly, for even $r$, we have,
\begin{align*}
    L_2(r)
    &=
    \sum_{j=0}^{\frac{r}{2}-1}
    \frac{K+r-2j-2}{2(K-1)} +
    \sum_{j=\frac{r}{2}}^{r-2}
    \frac{K-r+2(j+1)}{2(K-1)} \\
    &=
    \sum_{j=0}^{\frac{r}{2}-2}
    \frac{K+r-2j-2}{2(K-1)}  +
    \frac{
    K+r-2\left(\frac{r}{2}-1\right)-2
    }{
    2(K-1)
    } \\
    &\quad +
    \sum_{j=\frac{r}{2}}^{r-2}
    \frac{K-r+2(j+1)}{2(K-1)} \\
    &\stackrel{(a)}{=}
    \sum_{j=0}^{\frac{r}{2}-2}
    \frac{K+r-2j-2}{2(K-1)}  +
    \frac{K}{2(K-1)} \\
    &\quad +
    \sum_{j'=0}^{\frac{r}{2}-2}
    \frac{K-r+2(r-2-j'+1)}{2(K-1)} \\
    &=
    2\sum_{j=0}^{\frac{r}{2}-2}
    \frac{K+r-2j-2}{2(K-1)}
    +
    \frac{K}{2(K-1)} \\
    &=
    \left(
    \frac{2}{2(K-1)}
    \right)
    \Bigg[
    \left(\frac{r}{2}-1\right)(K+r-2)
    \\
    &\qquad -
    \left(\frac{r}{2}-2\right)
    \left(\frac{r}{2}-1\right)
    \Bigg]
    +
    \frac{K}{2(K-1)} \\
    &=
    \left(
    \frac{1}{2(K-1)}
    \right)
    (r-2)
    \left(
    K+\frac{r}{2}
    \right)
    +
    \frac{K}{2(K-1)},
\end{align*}
where $(a)$ is obtained by changing the variable $j' = (r-2)-j$.

The total communication load is therefore $\Lrem(r)=\frac{K-r}{(K-1)} + \min(L_1(r),L_2(r))$. 

\subsection{Advantage of Algorithm \ref{algo:rebalancing} over the uncoded scheme}
\label{subsec:advantage}
In this subsection, we bound the advantage of the rebalancing schemes presented in this work, over the uncoded scheme, in which the nodes simply exchange all the data which was available at the removed node via uncoded transmissions. We know that the load of the uncoded scheme is $L_u(r)=\frac{rT}{T}=r$. Consider the ratio of the communication load of Scheme 1 to that of the uncoded scheme. We then have the following sequence of equations. 
    \begin{align*}
        \frac{\frac{K-r}{K-1} + L_1(r)}{L_u(r)} &= \frac{1}{r} \left( \frac{K-r}{K-1} + \frac{(K-r)(2r-1)}{K-1} \right) \\
        &= \frac{K-r}{r(K-1)} \left(1 + 2r -1\right) \\
        &= \frac{2(K-r)}{K-1} = 2 \left( \frac{(K-1)-(r-1)}{K-1} \right) \\
        &= 2\left(1 - \frac{r-1}{K-1} \right).
    \end{align*}
    Now we do the same for Scheme 2.
\begin{align*}
    &\frac{\frac{K-r}{K-1} + L_2(r)}{L_u(r)}\\
    &=
    \frac{1}{r}
    \Bigg(
    \frac{K-r}{K-1}
    +
    \frac{r-1}{2(K-1)}
    \left(
    K+\frac{r-1}{2}
    \right)
    \Bigg) \\
    &=
    \frac{1}{2r(K-1)}
    \Bigg(
    2K-2r+rK-K  +
    \frac{(r-1)^2}{2}
    \Bigg) \\
    &=
    \frac{1}{2r(K-1)}
    \Bigg(
    K(r+1)
    +
    \frac{(r-1)^2-4r}{2}
    \Bigg) \\
    &<
    \frac{1}{2r(K-1)}
    \Bigg(
    K(r+1)
    +
    \frac{r^2-4r}{2}
    \Bigg) \\
    &\leq
    \frac{(K-1+1)(r+1)}
    {2r(K-1)}
    +
    \frac{r-4}{4(K-1)} \\
    &\leq
    \frac{1}{2}
    +
    \frac{r+K}{2r(K-1)}
    +
    \frac{r-4}{4(K-1)} \\
    &\leq
    \frac{1}{2}
    +
    \frac{1}{2(K-1)}
    \left(
    \frac{K-r}{r}
    +
    \frac{r}{2}
    \right) \\
    &<
    \frac{1}{2}
    +
    \frac{1}{2r}
    +
    \frac{r}{4(K-1)},
\end{align*}
    where the last inequality follows by using the fact that $K-r\leq K-1$. Observe that $\frac{1}{2} + \frac{1}{2r}+\frac{r}{4(K-1)}<1$, as $r\geq 3$ and $K-1\geq r$. 
    As the choice of the scheme selected for transmissions is based on the minimum load of Scheme 1 and Scheme 2, we have the result in the theorem. This completes the proof of the node-removal part of Theorem \ref{thm:achieve}.

\section{Proof of Theorem \ref{thm:achieve}:  Rebalancing for single node addition in cyclic databases}
\label{sec:rebalancing_nodeaddition}
We consider the case of a $r$-balanced cyclic database system when a new node is added. Let this new empty node be indexed by $K+1$. For this imbalance situation, we present a rebalancing algorithm (Algorithm \ref{algo:nodeaddition}) in which nodes split the existing data segments and broadcast appropriate subsegments, so that the target database which is a $r$-balanced cyclic database on $K+1$ nodes, can be achieved. Each subsegment's size is assumed to be an integral multiple of $\frac{1}{K+1}$. This is without loss of generality, by the condition on the size $T$ of each segment as in Theorem \ref{thm:achieve}. Since node $K+1$ starts empty, there are no coding opportunities; hence, the rebalancing scheme uses only uncoded transmissions. We show that this rebalancing scheme achieves a normalized communication load of $\frac{r}{K+1}$, which is known to be optimal from the results of \cite{krishnan2020coded}. This proves the node-addition part of Theorem \ref{thm:achieve}. 

\begin{algorithm}[H]
\caption{Rebalancing Scheme for Single Node Addition}
\label{algo:nodeaddition}
\begin{algorithmic}[1]
\Procedure{Delivery Scheme}{}
    \For{each $i \in [K]$}
        \State Split $W_i$ into two subsegments, labelled $\fW_i$ of size $\frac{K}{K+1}$ and $W_i^{\{K+1\} \cup [\min(r-1,i-1)]}$ of size $\frac{1}{K+1}$. 
        \State Node $i$ broadcasts $W_i^{\{K+1\} \cup [\min(r-1,i-1)]}$.
    \EndFor
    \State Each node in $\{(K+1) \boxplus_{K+1} \langle r \rangle\}$ initializes the segment $\fW_{K+1}$ as an empty vector.
    \For{each $i \in [K]$}
        \State Each node in $\{(K+1) \boxplus_{K+1} \langle r \rangle\}$ performs the concatenation $\fW_{K+1} = \fW_{K+1} | W_i^{\{K+1\}\cup [\min(r-1,i-1)]}$.
    \EndFor
    \For{each $i=(K-r+2)$ to $K$}
        \State Node $i$ transmits $\fW_{i}$ to node $K+1$.    
    \EndFor
    \For{each $i=1$ to $r-1$}
        \State Node $i$ discards $\fW_{K-r+1+i}$.
    \EndFor
\EndProcedure

\end{algorithmic}
\end{algorithm}

\begin{remark}
\label{rem:computationalcomplexity-nodeadd}
    The computational complexity of Algorithm \ref{algo:nodeaddition} is shown to be $O(KT)$ in Appendix \ref{app:computationcomplexitynodeaddition}. 
\end{remark}

\textbf{Note:} Once the target segments $\fW_1, \dots, \fW_{K+1}$ are recovered at the required nodes, any extra bits present at the node are discarded.

\subsection{Correctness of Algorithm \ref{algo:nodeaddition}}
We verify the correctness of the rebalancing algorithm, i.e., we check that the target $r$-balanced cyclic database on $K+1$ nodes is achieved post-rebalancing. To achieve the target database, each target segment $\fW_i$, for $i \in [K+1]$, must be of size $\frac{K}{K+1}$ and stored exactly in $i \boxplus_{K+1} \langle r\rangle$. Consider the segments $\fW_{i}$ for each $i \in [K-r+1]$. Since these were a part of $W_i$, they are already present exactly at nodes $\{i \boxplus_{K+1} \langle r\rangle \}$. 

Now, consider the segments $\fW_i:i \in \{(K-r+1) \boxplus_{K} \langle r \rangle\}$. We recall $S_i$ as the set of nodes where $W_i$ is present in the initial database. Let $\fS_i$ be the set of nodes where it must be present in the final database. Now, the nodes where $\fW_i$ is not present and must be delivered to are given by $\fS_i \backslash S_i = \{K+1\}$. This is performed in lines 10-12 of Algorithm \ref{algo:nodeaddition}. Also, the nodes where $W_i$ is present but $\fW_i$ must not be present are given by $S_i \backslash \fS_i = \{i-K+r-1\}$. This node discards $\fW_i$ in lines 13-15 of Algorithm \ref{algo:nodeaddition}. Finally, $\fW_{K+1}$ must be present in nodes $\{(K+1) \boxplus_{K+1} \langle r \rangle\}$. This can be obtained by the those nodes from the broadcasts made in line 4 in Algorithm \ref{algo:nodeaddition} and the concatenation performed in lines 7-9.

Finally, It is easy to calculate that each target segment $\fW_i:i \in [K+1]$ is of size $\frac{K}{K+1}$. Thus, the final database satisfies the cyclic storage condition.

\subsection{Communication Load of Algorithm \ref{algo:nodeaddition}}
For broadcasting each $W_i^{\{K+1\} \cup [\min(r-1,i-1)]}$ of size $\frac{1}{K+1}$, $\forall i \in [K]$, the communication load incurred is $K.\frac{1}{K+1} = \frac{K}{K+1}$. Now, to transmit $\fW_{K-r+2} \dots \fW_{K}$ to node $K+1$, the communication load incurred is $(r-1).\frac{K}{K+1}$. Hence, the total communication load is $\Ladd(r) = \frac{rK}{K+1}$, which is known to be optimal from \cite{krishnan2020coded}. 



\section{Lower Bound for Rebalancing Load for Node-Removal}
\label{sec:lower_bound_revised}

In \cite{krishnan2020coded}, a lower bound on the rebalancing load was obtained, which is applicable for any $r$-balanced database (not necessarily cyclic), which is $\frac{r}{r-1}$. In this section, we obtain a new, improved lower bound under the constraint that the initial and the final databases are both cyclic. Subsection \ref{subsec:lowerboundmaintheorem_revised}, presents this lower bound, obtained via a careful analysis of the required quantum of data to be delivered to each node and its relationship to the data placement in the initial database. In Subsection \ref{subsec:multiGap_revised}, we show that the achievable loads of Theorem \ref{thm:achieve} are within a multiplicative gap from the lower bound shown in Subsection \ref{subsec:lowerboundmaintheorem_revised}, thus proving part $2)$ of Theorem \ref{thm:achieve}.

We recall the relationship of the rebalancing phase with \textit{index coding}, introduced by Birk and Kol \cite{ISCOD_BiK}. When the demands of the nodes are fixed, the data exchange problem involved in the transmission phase of the rebalancing algorithm becomes an index coding problem. Any such index coding problem can be represented \cite{neely2013dynamic} using a corresponding directed bipartite graph $\cal{G}$ with left-vertices as the nodes and right vertices as the demanded subsegments. The definition of the edges in this graph is as follows. 
\begin{itemize}
    \item For each node $i$, a single solid edge points from the vertex corresponding to the demanded subsegment by node $i$ to the vertex corresponding to node $i$. 
    \item For each node $i$, dashed edges point from the vertex corresponding to node $i$ to the vertices corresponding to the subsegments that are present in node $i$. 
\end{itemize}
An information-theoretic lower bound for an index coding problem (which is applicable even if there is a central server node, i.e. one having all the data) was then derived in \cite{neely2013dynamic} by extracting an \textit{acyclic induced subgraph}\footnote{An acyclic induced subgraph of a graph $\cal{G}$ is an induced subgraph of $\cal{G}$ that is not cyclic.} from $\cal{G}$ by performing some pruning operations. We use the well-known maximum acyclic induced subgraph (MAIS) lower bound for index coding that can be obtained from the results of \cite{neely2013dynamic}, which we summarize below, for our context.

\begin{definition}[Maximum Acyclic Induced Subgraph (MAIS)]
An acyclic induced subgraph of the index coding graph $\mathcal{G}$ is an 
induced subgraph of $\mathcal{G}$ that contains no directed cycle. It is 
composed of a collection of subsegments $B_{k_1},\hdots, B_{k_l}$ and 
$l$ nodes $k_1,\hdots,k_l$, such that for each $i\in[l]$, $B_{k_i}$ is 
demanded at node $k_i$, and $B_{k_i}$ is not available as side-information 
at nodes $k_i, k_{i-1},\hdots,k_1$. The Maximum Acyclic Induced Subgraph (MAIS) is the acyclic induced subgraph for which $|\cup_{i=1}^l B_{k_i}|$ 
is maximum.
\end{definition}

\begin{lemma}\cite[Lemma 1]{neely2013dynamic}
\label{obs:structureofacyclic}
For any index coding problem represented by graph $\mathcal{G}$, the number 
of bits to be communicated is at least $|\cup_{i=1}^l B_{k_i}|$ for a 
MAIS with nodes $k_1,\hdots,k_l$ and corresponding 
demands $B_{k_1},\hdots,B_{k_l}$.
\end{lemma}


The key differences between the rebalancing problem and the index coding problem are that (a) the demands are not specified in the rebalancing problem, and (b) there is no central server which has access to all symbols. Firstly, as every central-server-based bound is still applicable when there is no such node available, we can apply the index coding bound from \cite{neely2013dynamic} for our purpose as well. Further, when the initial and final databases are constrained to be cyclic, then we show that the quantum of data to be placed in each of the `survivor' nodes (the nodes not removed) must satisfy some corresponding lower bound, and further this minimum quantum of data must be pre-available only in some specific subset of other survivor nodes. Using these observations, and the acyclic-subgraph technique from \cite{neely2013dynamic}, we find a lower bound for the rebalancing problem, when the target database is cyclic.   

\subsection{A Lower Bound on the Load via Acyclic Subgraphs}
\label{subsec:lowerboundmaintheorem_revised}
In this subsection, we prove some structural lemmas regarding the index coding graph associated with the rebalancing problem for any choice of the demands (the demand at any node is the data to be delivered to it). We make the assumption that the demand at each survivor node was present in the removed node (which we assume throughout is node $K$). Note that copies of the data in node $K$ are in fact required to be moved to new nodes in any rebalancing scheme, to reinstate the replication factor. That all demands are composed of the data in node $K$ is therefore a natural assumption.

We split our analysis into two scenarios, based on the condition that $r < (K+1)/2$ (the `low-$r$' regime) or otherwise (the `high-$r$' regime). We do this, as in the low-$r$ regime, complicated `wrap-around's in the placement do not occur, whereas in the high-$r$ regime, they do occur. This leads to different bounds in these regimes. 

\subsubsection{Case 1: $r< (K+1)/2$ (low-$r$ regime)}

Towards our analysis, we first make the following observations, which are straightforward to prove.  
\begin{observation}
\label{obs:commonstorageconsecutivenodes}
    In any $r$-balanced cyclic database with $K$ nodes, the number of common segments  in any $r$ consecutive nodes $\{j\boxplus_K\langle r\rangle\}$ is precisely $1$, which is the segment $W_j$. Further, any $r+1$ consecutive nodes $\{j\boxplus_K\langle r+1\rangle\}$ do not share any common segment. 
\end{observation}
\begin{observation}
\label{obs:distinctstorageconsecutivenodes}
In any $r$-balanced cyclic database with $K$ nodes, the number of distinct segments in any consecutive $l$ nodes is $\min(K,r+l-1)$.     
\end{observation}
The following lemma shows that there are two acyclic induced subgraphs of the index coding graph associated with the rebalancing problem, given any choices for the demands at the nodes. 

\begin{lemma}
\label{lemma:lowracyclicsubgraphs}
    Let the demand at node $k\in[K-1]$ be $B_k$ in the target database of an arbitrary feasible rebalancing scheme, as per Definition \ref{defn:rebalancing}. For $r<(K+1)/2$, each of the following two collections of nodes (along with their demands), defines an acyclic induced subgraph of the associated index coding graph.
    \begin{itemize}
        \item The nodes $N_1\triangleq \{1,\hdots,r-1\}$ and the nodes $N_3\triangleq \{r,\hdots,K-r\}$. 
        \item The nodes $N_2\triangleq \{K-r+1,K-r+2,\hdots,K-1\}$ and the nodes $N_3= \{r,\hdots,K-r\}$. 
    \end{itemize}
\end{lemma}
\begin{IEEEproof}
The segments in the removed node $K$ are precisely $D_K=\{W_K,W_{K-1},\hdots,W_{K-r+1}\}$. Thus, the existing contents of the nodes $D_k:k\in N_1$ and $D_k:k\in N_2$ satisfy the following conditions.

\begin{itemize}
    \item $D_{k}\cap D_K=\{W_{K-r+1+k},W_{K-r+k+2},\hdots,W_K\}$, for $k\in N_1$.
    \item $D_{k}\cap D_K=\{W_{K-r+1},W_{K-r+2},\hdots,W_k\}$, for $k\in N_2$. 
    \end{itemize}
We thus note that $D_{k}\cap D_K\subset D_{k-1}$ for $k\in\{2,\hdots,r-1\}$, and $D_{k}\cap D_K\subset D_{k+1}$
for $k\in\{K-2,\hdots,K-r+1\}$. 

Note that the demand at each node $k\in N_1\cup N_2$ satisfies $B_k\subseteq D_K\setminus D_k$. By the structure of $D_k$ observed above, we see that $B_k\cap(\cup_{j\in\{k,k+1,\hdots,r-1\}}D_j)=\emptyset$, for $k\in N_1$, i.e., the demand of node $k$ is unavailable in the nodes $\{k,\hdots,r-1\}$. Similarly, for nodes $k\in N_2$ the demand of the node $k$ is unavailable in the nodes $\{k,k-1,k-2,\hdots,K-r+1\}$, and thus  $B_k\cap(\cup_{j\in\{k,k-1,\hdots,K-r+1\}}D_j)=\emptyset$. By Lemma \ref{obs:structureofacyclic}, we see that the nodes $N_1$ along with their demands form an acyclic induced subgraph. Similarly, the nodes $N_2$ along with their demands form another acyclic induced subgraph. 

Now, the survivor nodes apart from the nodes $N_1\cup N_2$ are precisely the nodes $N_3=\{r,r+1,\hdots,K-r\}$. The content in these nodes $k\in N_3$, is given by $D_k=\{k\boxminus_K\langle r\rangle\}$. From this, it is easy to observe that $D_k\cap D_K=\phi$, for all $k\in N_3$. Thus, the nodes $k\in N_3$ do not contain any bit of the demands $B_k:k\in[K-1]$. With this and prior analysis, we obtain the proof of the claim. 
\end{IEEEproof}

Lemma \ref{lemma:lowracyclicsubgraphs} immediately gives us two lower bounds based on the respective acyclic induced subgraphs. What remains to be done is to bound the number of bits in the demands $B_k:k\in[K-1]$. A natural assumption would that $|B_k|={rT}/(K-1)$, as it must be that $\cup_{k\in[K-1]}B_k=D_K$ and $|D_K|=rT$. However, it turns out that this is based on the assumption that the demands are all disjoint subsets, and also does not take into account the cyclic structure of the target database. The following lemma shows that, accounting for the cyclicity, we can obtain better (larger) bounds on sizes of the node-demands. 

\begin{lemma}
\label{lemma:lowrdemandslowerbounds}
    Suppose $r<(K+1)/2$. Then, in any feasible rebalancing scheme with a target cyclic $r$-balanced database on $K-1$ nodes, the demands $B_k:k\in[K-1]$ must contain respective subsets $B_k':k\in[K-1]$ that satisfy the following constraints. 
    \begin{enumerate}
        \item $B'_{k_1}\cap B_{k_2}'=\emptyset$, for all distinct $k_1,k_2 \in[K-1]$. 
        \item $|B'_j|+|B'_{K-r+j}|\geq \frac{KT}{K-1}$, for all $j\in N_1=[r-1]$.
        \item $|\cup_{k\in N_3}B'_k|\geq \frac{(K-r)T}{K-1}$.
    \end{enumerate}
\end{lemma}

\begin{IEEEproof} Let the new segments in the target cyclic database be denoted as $\fW_j:j\in[K-1]$. We observe that each new segment must be of size $T'=KT/(K-1)$, where $T$ is the size of the segments in the original database. We also recall the definition of the nodes $N_1,N_3$ and $N_2$ in Lemma \ref{lemma:lowracyclicsubgraphs}.

We first prove \textit{3)}. By Observation \ref{obs:distinctstorageconsecutivenodes}, the collection of $K-2r+1$ nodes in $N_3$ must contain $\min(K-1,r+(K-2r+1)-1)=K-r$ distinct new segments (of size $T'$ each), post rebalancing. Now, before rebalancing, these nodes contain $K-r$ distinct segments of size $T$. Thus, the number of distinct bits that needs to be delivered to the nodes in $N_3$ is at least $(K-r)(T'-T)=\frac{(K-r)T}{K-1}$. We partition this collection of distinct bits to be delivered into disjoint parts $B'_k:k\in N_3$, where $B'_k\subset B_k$ must be delivered to node $k\in N_3$. Thus, we see that the disjoint union $\cup_{k\in N_3} B'_k$ satisfies \textit{3)}.

We now prove \textit{1) and 2)}. Observe that, for each $j\in N_1$ the node $K-r+j$ lies in $N_2$. We first seek to analyze how many common segments lie in this pair of nodes $j,K-r+j$ in the initial database, and how many common new segments they will share in the target database. This will help us determine the number of bits that are to be delivered in total to this pair of nodes. Note that the number of common segments depend on how `far' the node $j$ and $K-r+j$ are from each other, as per the note in Observation \ref{obs:commonstorageconsecutivenodes}. In other words, we want to find out how many indices we have to shift `up', starting from $j$ to reach $K-r+j$ and vice-versa as well. 

To do this, we note that $\{(K-r+j)\boxplus_K\langle r+1 \rangle\}=\{K-r+j,K-r+j+1,\hdots,K,1,\hdots,j\}$. Further, the set $\{j\boxplus_{K}\langle r \rangle\}=\{j,\hdots,j+(r-1)\}$ does not contain the node $K-r+j$ as $j+r-1<K-r+j$ (true, since we have $2r< K+1$ in this scenario). Thus, in the original database, the nodes $j\in N_1$ and $K-r+j\in N_2$ share no common segment. Now, post-rebalancing, we have the relationship, $\{(K-r+j)\boxplus_{K-1}\langle r \rangle\}=\{K-r+j,\hdots,K-1,1,\hdots,j\}$. Further, $\{j\boxplus_{K-1}\langle r \rangle\}=\{j,\hdots,j+(r-1)\}$ again does not contain $K-r+j$, for any $j\in N_1$. Thus, we note that, post-rebalancing, the nodes $K-r+j\in N_2$ and $j\in N_1$ must share precisely one new segment, which is of size $T'=\frac{KT}{K-1}$. This new segment must indeed be $\fW_{K-r+j}$, by Observation \ref{obs:commonstorageconsecutivenodes}.    


Now, let $B'_j\triangleq \fW_{K-r+j}\cap B_j$ and $B'_{K-r+j}=(\fW_{K-r+j}\cap B_{K-r+j})\setminus B_j$. By our prior arguments, we see that $\fW_{K-r+j}=B'_j \cup B'_{K-r+j}$, and by definition this is a disjoint union. Thus, $|B'_j|+|B'_{K-r+j}|= \frac{KT}{K-1}$. Further, note that, as we move through $j\in[r-1]$, the new segments $\fW_{K-r+j}$ must also be distinct and therefore be comprised of disjoint collections of bits. Thus, we see that, all the parts $B'_j$ and $B'_{K-r+j}$, for $j\in N_1$ are all disjoint. Further, none of these parts are to be placed in the nodes $\{r,\hdots,K-r\}$, since the new segment $\fW_{K-r+j}$ is intended only for the nodes $\{K-r+j,\hdots,K-1,1,\hdots,j\}$. Thus,  the parts $B'_j$ and $B'_{K-r+j}$, for $j\in N_1$ must be disjoint from the demands of the nodes $\{r,\hdots,K-r\}$. This completes the proof.

\end{IEEEproof}

\begin{figure}[ht]
    \centering
    \begin{subfigure}[ht]{0.48\linewidth}
        \centering
        \includegraphics[width=\linewidth]{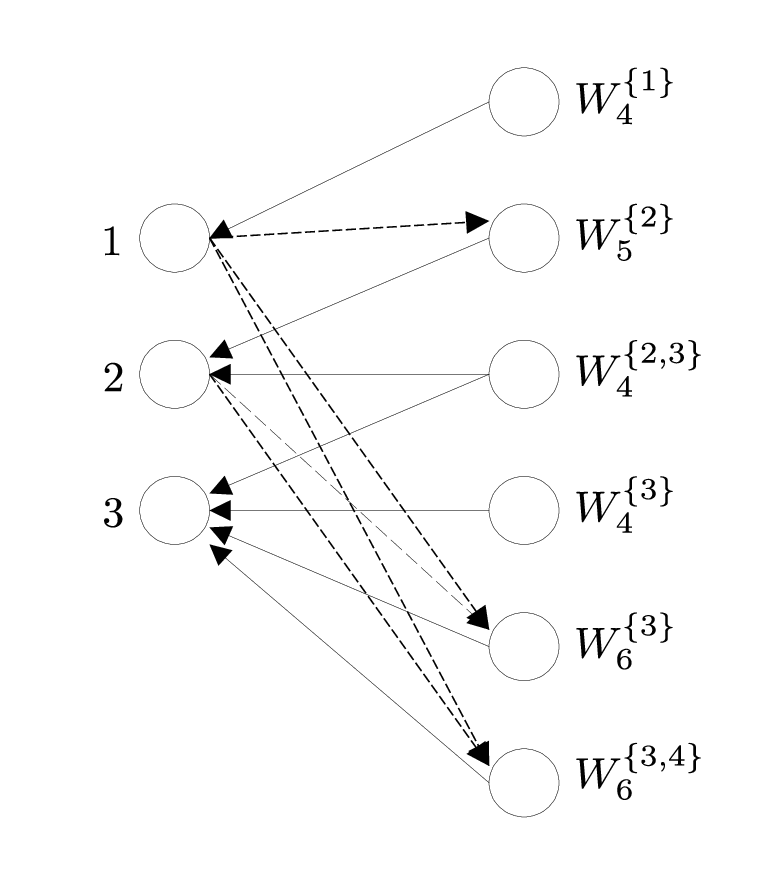}
        \caption{Subgraph corresponding to nodes in $N_1 \cup N_3$ and their demands, for $K=6, r=3$.}
        \label{fig:example1a}
    \end{subfigure}
    \hfill
    \begin{subfigure}[ht]{0.48\linewidth}
        \centering
        \includegraphics[width=\linewidth]{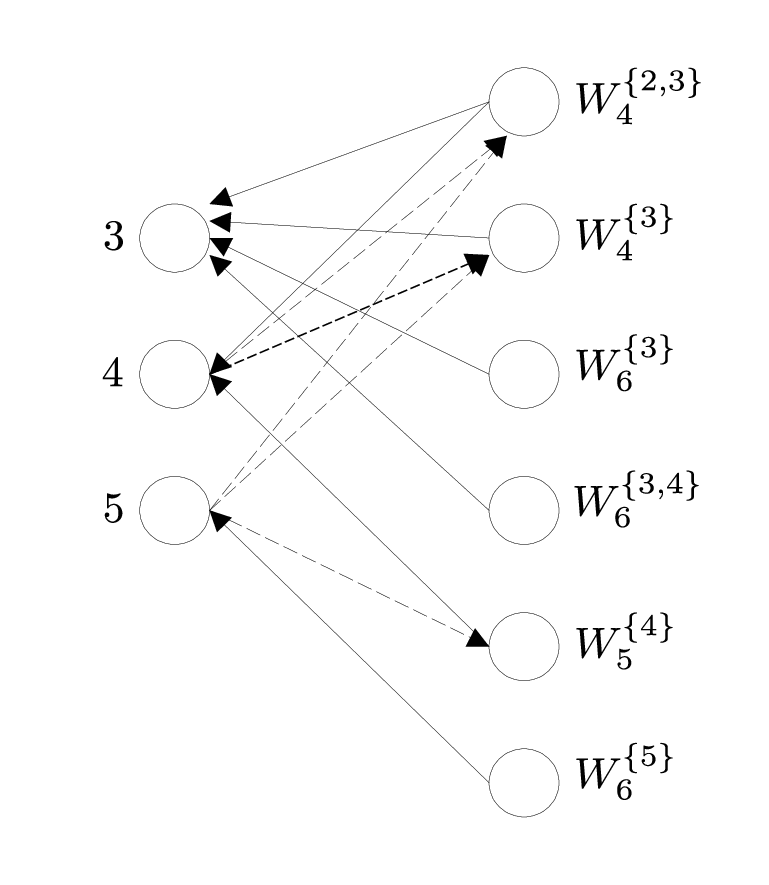}
        \caption{Subgraph corresponding to nodes in $N_2 \cup N_3$ and their demands, for $K=6, r=3$.}
        \label{fig:example1b}
    \end{subfigure}
    \caption{Subgraphs illustrating Lemma \ref{lemma:lowracyclicsubgraphs}, for $K=6, r=3$.}
    \label{fig:example1}
\end{figure}

\textbf{Example illustrating Lemma \ref{lemma:lowracyclicsubgraphs} and Lemma \ref{lemma:lowrdemandslowerbounds}:} We consider the database with $K = 6$ nodes and replication factor $r = 3$ from Section \ref{sub:example}. Here, we have $r = 3 < \frac{6+1}{2}$. From Lemma \ref{lemma:lowracyclicsubgraphs}, we have $N_1 = \{1,2\}, N_2 = \{4,5\}$, and $N_3 = \{3\}$. In Table \ref{tab:demands_example1}, the demanded subsegments and their respective sizes, for each node, are as per Algorithm \ref{algo:split}. Figure~\ref{fig:example1} illustrates the two subgraphs described in Lemma~\ref{lemma:lowracyclicsubgraphs}.


\begin{table*}[ht]
\centering
\begin{tabular}{c l l c}
\hline
Node $k$ & Demands & Disjoint subset $B_k'$ & Size of $B_k'$ \\
\hline
1 & $\{ W_4^{\{1\}} \}$ 
  & $\{ W_4^{\{1\}} \}$ 
  & $\frac{7T}{10}$ \\

2 & $\{ W_5^{\{2\}},\; W_4^{\{2,3\}} \}$ 
  & $\{ W_5^{\{2\}} \}$ 
  & $\frac{5T}{10}$ \\

3 & $\{ W_4^{\{2,3\}},\; W_4^{\{3\}},\; W_6^{\{3\}},\; W_6^{\{3,4\}} \}$ 
  & $\{ W_4^{\{2,3\}},\; W_4^{\{3\}},\; W_6^{\{3\}},\; W_6^{\{3,4\}} \}$ 
  & $\frac{6T}{10}$ \\

4 & $\{ W_6^{\{3,4\}},\; W_5^{\{4\}} \}$ 
  & $\{W_5^{\{4\}} \}$ 
  & $\frac{5T}{10}$ \\

5 & $\{ W_6^{\{5\}} \}$ 
  & $\{ W_6^{\{5\}} \}$ 
  & $\frac{7T}{10}$ \\
\hline
\end{tabular}
\caption{Demanded subsegments, corresponding disjoint subsets $B_k'$ and their sizes for the nodes in $[K-1]$, for $K=6,r=3$.}
\label{tab:demands_example1}
\end{table*}

As we can see from Fig. \ref{fig:example1a} and \ref{fig:example1b}, the subgraphs are acyclic. Moreover, the sum of the sizes of the subsegments in both the graphs is $\frac{18T}{10}$, which matches the bound given in Theorem \ref{theorem:lowr-lowerboundexpression}. It is also easy to see from Table \ref{tab:demands_example1} that all the parts of Lemma \ref{lemma:lowrdemandslowerbounds} hold.

Using Lemmas \ref{lemma:lowracyclicsubgraphs} and \ref{lemma:lowrdemandslowerbounds}, and the existing lower bound from \cite{krishnan2020coded}, we obtain the following lower bound on the optimal rebalancing load, for the low-$r$ regime. 
\begin{theorem}
\label{theorem:lowr-lowerboundexpression}
Let $r< (K+1)/2$. The optimal rebalancing load $\Lrem^*(r)$ for the node removal scenario, satisfies the following lower bound, when the target database is cyclic. 
\begin{align}
\label{eqn:lowerboundlow-r-regime}
    \Lrem^*(r)\geq 
        \max\left(\frac{(K-r)}{K-1}+\frac{K(r-1)}{2(K-1)},\frac{r}{r-1}\right).
\end{align}    
Further, the following holds true.
\begin{align}
\label{eqn:firstmaxlowr-regime}
&\max\Bigg(
\frac{K-r}{K-1}
+
\frac{K(r-1)}{2(K-1)},
\,
\frac{r}{r-1}
\Bigg)
\nonumber \\
&\qquad =
\frac{K-r}{K-1}
+
\frac{K(r-1)}{2(K-1)},
\forall r \geq 3,\;
\forall K \geq r+1.
\end{align}
\end{theorem}
\begin{IEEEproof}
We first prove \eqref{eqn:lowerboundlow-r-regime}. Firstly, recall that the $r/(r-1)$ lower bound is applicable to all target databases, as per the result in \cite{krishnan2020coded}. From Lemmas \ref{obs:structureofacyclic} and  \ref{lemma:lowracyclicsubgraphs}, and Lemma \ref{lemma:lowrdemandslowerbounds} (part \textit{1)}), the load $\Lrem(r)$ of any rebalancing scheme must satisfy the following bounds.
\begin{align}
    \label{eqn:lowerbound1}
    \Lrem(r)\geq \frac{1}{T}|\cup_{k\in N_1\cup N_3}B'_k|\geq \frac{1}{T}\left(\sum_{k\in N_1}|B'_k|+|\cup_{k\in N_3}B'_k|\right),\\
    \label{eqn:lowerbound2}
    \Lrem(r)\geq \frac{1}{T}|\cup_{k\in N_2\cup N_3}B'_k|\geq \frac{1}{T}\left(\sum_{k\in N_2}|B'_k|+|\cup_{k\in N_3}B'_k|\right).
\end{align}
Adding \eqref{eqn:lowerbound1} and \eqref{eqn:lowerbound2}, we get
\begin{align*}
    2\Lrem(r)&\geq \frac{1}{T}\left(\sum_{k\in N_1}|B'_k|+\sum_{k\in N_2}|B'_k|+2|\cup_{k\in N_3}B'_k|\right)\\
    &=\frac{1}{T}\left(\sum_{k\in N_1}(|B'_k|+|B'_{K-r+k}|)+2|\cup_{k\in N_3}B'_k|\right)
\end{align*}
Now, using Lemma \ref{lemma:lowrdemandslowerbounds} (parts \textit{2)} and \textit{3)}) and since $|N_1|=r-1$, we get
\begin{align*}
    \Lrem(r)&\geq \frac{1}{2}(r-1)\frac{K}{K-1}+\frac{K-r}{K-1},
\end{align*}
which completes the proof, as this holds for any rebalancing scheme with the target database being cyclic.

Now, for the next part of the proof, let $f(K) = \frac{(K-r)}{K-1}+\frac{K(r-1)}{2(K-1)}$. On simplifying, we get $f(K) = \frac{K+r(K-2)}{2(K-1)}$. To prove \eqref{eqn:firstmaxlowr-regime}, we need to show that $f(K) \geq \frac{r}{r-1}, \forall r \geq 3, \forall K\geq r+1.$
Consider the derivative $f'(K)$ with respect to $K$,
\begin{align*}
    f'(K) &= \frac{2(K-1)(1+r)-2(K+r(K-2))}{4(K-1)^2} \\
    &= \frac{r-1}{2(K-1)^2}.
\end{align*}
Thus, $f$ is increasing in $K$, implying that it suffices to check the inequality at $K = r+1$. Now, $f(r+1) = \frac{r^2+1}{2r}$. Therefore, we have
\begin{align*}
    \frac{r^2+1}{2r} - \frac{r}{r-1} &\geq \frac{(r^2+1)(r-1) -2r^2}{2r(r-1)}\\ 
    &\geq \frac{r^2(r-3)+r-1}{2r(r-1)} \\ 
    &\geq 0,
\end{align*}
which is true for $r \geq 3$. Thus, we have proved \eqref{eqn:firstmaxlowr-regime}. This completes the proof. 
\end{IEEEproof}

\subsubsection{Case 2: $r\geq (K+1)/2$ (high-$r$ regime)}
We now turn to the high-$r$ regime. We provide the equivalents of Lemmas \ref{lemma:lowracyclicsubgraphs} and \ref{lemma:lowrdemandslowerbounds} for this case, and then the lower bound for this case as well. As the ideas for the proofs are essentially similar to the low-$r$ regime, we will be brief, and mainly highlight the differences. 

The following lemma provides the two acyclic subgraphs of the index coding graph associated with the rebalancing problem. The key difference from Lemma \ref{lemma:lowracyclicsubgraphs} is that, unlike the low-$r$ regime, we can only take upto $K-r$ consecutive nodes on either side of the removed node $K$, to define our acyclic induced subgraph. Further, the `in-between' nodes that played a role in Lemma \ref{lemma:lowracyclicsubgraphs} (nodes $N_3$) are absent here. This is because of the fact that $r$ is large, which results in the breakdown of acyclicity, when we consider more than $K-r$ nodes on either side. 
\begin{lemma}
\label{lemma:highracyclicsubgraphs}
    Let the demand at node $k\in[K-1]$ be $B_k$ in the target database of an arbitrary feasible rebalancing scheme, as per Definition \ref{defn:rebalancing}. For $r\geq\frac{K+1}{2}$, each of the following two collections of nodes (along with their demands) defines an acyclic induced subgraph of the associated index coding graph.
    \begin{itemize}
        \item The nodes $N'_1\triangleq \{1,\hdots,K-r\}$.
        \item The nodes $N'_2\triangleq \{r,r+1,\hdots,K-1\}$. 
    \end{itemize}
\end{lemma}
\begin{IEEEproof}
The existing contents of the nodes $D_k:k\in N'_1\cup N'_2$ satisfy the following conditions.

\begin{itemize}
    \item $D_{k}\cap D_K=\{W_{K-r+1+k},W_{K-r+k+2},\hdots,W_K\}$, for $k\in N'_1$.
    \item $D_{k}\cap D_K=\{W_{K-r+1},W_{K-r+2},\hdots,W_k\}$, for $k\in N'_2$. 
    \end{itemize}
The remaining arguments of this proof proceed identically to the proof of Lemma \ref{lemma:lowracyclicsubgraphs}. 
\end{IEEEproof}
We now obtain lower bounds on the number of bits demanded by the nodes in $N'_1$ and $N'_2$. 

\begin{lemma}
\label{lemma:highrdemandslowerbounds}
    Suppose $r\geq\frac{K+1}{2}$. Then, in any feasible rebalancing scheme with a target cyclic $r$-balanced database on $K-1$ nodes, the demands $B_k:k\in N'_1\cup N'_2$ must contain subsets $B_k':k\in N'_1\cup N'_2$ that satisfy the following constraints. 
    \begin{enumerate}
        \item $B'_{k_1}\cap B'_{k_2} = \emptyset$, $\forall k_1 \neq k_2 \in N'_1$, and $\forall k_1 \neq k_2 \in N'_2$. 
        \item $|B'_j|+|B'_{j+(r-1)}|\geq \frac{2rT}{K-1}$, for all $j\in N'_1$. 
    \end{enumerate}
\end{lemma}
\begin{IEEEproof}
Firstly, we show that the nodes $j\in N'_1$ and $(r-1)+j\in N'_2$ share $2r-K$ segments in the original database, and must share one additional new segment in the rebalanced database. We use similar `shift' based arguments, as before, to show this. 

Observe that $\{j+(r-1),j+r,\hdots,K,1,\hdots,j\}$ is a collection of $(K-r+2)$ consecutive nodes, where $K-r+2\leq r$ as per our scenario. Further, $\{j,j+1,\hdots,j+(r-1)\}$ is a collection of $r$ consecutive nodes. Now, from the structure of the cyclic database, any consecutive sequence of $l$ nodes $\{j'\boxplus_K\langle l\rangle\}$ has $r-l+1$ common segments, which are $W_k:k\in\{j'\boxminus_K\langle r-l+1\rangle\}$.  Thus, the collection of common segments in the nodes $j$ and $j+(r-1)$ are the $2r-K-1$ segments given by $W_k:k\in\{(j+(r-1))\boxminus_K\langle 2r-K-1\rangle\}$ as well as the segment $W_j$. Thus, in total $2r-K$ segments, amounting to $(2r-K)T$ bits, must be common between the nodes $j$ and $j+(r-1)$, for each $j\in N'_1$, in the initial database. 

Post-rebalancing, the nodes $\{j,j+(r-1)\}$ must  share the new segments $\fW_k:k\in\{(j+(r-1))\boxminus_{K-1}\langle 2r-K\rangle\}$, as $\{j+(r-1),j+r,\hdots,K-1,1,\hdots,j\}$ is a collection of $(K-r+1)$ consecutive nodes.  Further, $\{j,j+1,\hdots,j+(r-1)\}$ continues to be a $r$-consecutive collection, and thus the new segment $\fW_j$ must also be common between the nodes $j,j+(r-1)$ post-rebalancing. Thus, post-rebalancing, nodes $j,j+(r-1)$ must share a collection of $(2r-K+1)$ new segments. Since each target segment has size $\frac{KT}{K-1}$, the total amount of shared bits they must mutually possess is $(2r-K+1)\cdot\frac{KT}{K-1}$. The distinct new bits that this pair of nodes must collectively receive should be at least the difference between the shared target data they need and the shared original data they already possess. Subtracting the two quantities gives $(2r-K+1)\cdot\frac{KT}{K-1}$ - $(2r-K)T$, which simplifies to exactly $\frac{2rT}{K-1}$. Specifically, the demand of node $j$ should include $((\cup_{k\in\{(j+(r-1))\boxminus_{K-1}\langle 2r-K\rangle\}}\fW_k)\cup \fW_j)\setminus (D_j \setminus D_K)$ and that of node $j+(r-1)$ must include $((\cup_{k\in\{(j+(r-1))\boxminus_{K-1}\langle 2r-K\rangle\}}\fW_k)\cup \fW_j)\setminus (D_{j+(r-1)} \setminus D_K)$.

Further, these demands must be distinct across nodes in $N'_1$, as we run through all $j$. This is because, when $j$ increases to $j+1$, the segment $\fW_{j+1}$ will be present in the demands of node $j+1$ (while absent in demands of $j$). Similar observations can be made across nodes in $N'_2$, as we change $j$. This completes the proof.


\end{IEEEproof}

\begin{figure}[ht]
    \centering
    \begin{subfigure}[ht]{0.48\linewidth}
        \centering
        \includegraphics[width=\linewidth]{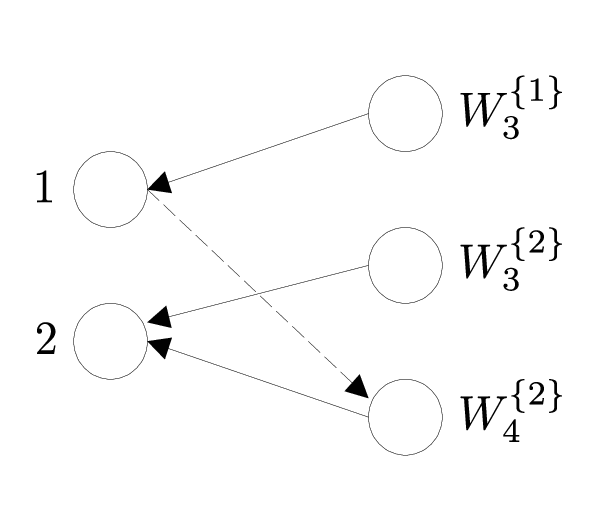}
        \caption{Subgraph corresponding to nodes in $N_1'$ and their demands, for $K=8, r=6$.}
        \label{fig:example2a}
    \end{subfigure}
    \hfill
    \begin{subfigure}[ht]{0.48\linewidth}
        \centering
        \includegraphics[width=\linewidth]{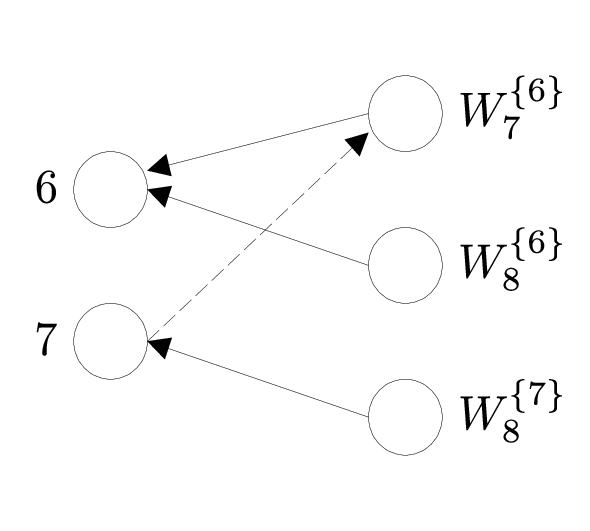}
        \caption{Subgraph corresponding to nodes in $N_2'$ and their demands, for $K=8, r=6$.}
        \label{fig:example2b}
    \end{subfigure}
    \caption{Subgraphs illustrating Lemma \ref{lemma:highracyclicsubgraphs}, for $K=8, r=6$.}
    \label{fig:example2}
\end{figure}

\textbf{Example illustrating Lemma \ref{lemma:highracyclicsubgraphs} and Lemma \ref{lemma:highrdemandslowerbounds}:} We consider the database with $K = 8$ nodes and replication factor $r = 6$ from Section \ref{sub:example}. Here, we have $r = 6 > \frac{8+1}{2}$. From Lemma \ref{lemma:highracyclicsubgraphs}, we have $N_1' = \{1,2\}$ and $N_2' = \{6,7\}$. In Table \ref{tab:demands_example2}, the demanded subsegments and their respective sizes, for each node, are as per Algorithm \ref{algo:split}. Figure~\ref{fig:example2} illustrates the two subgraphs described in Lemma~\ref{lemma:highracyclicsubgraphs}. 


\begin{table}[ht]
\centering
\begin{tabular}{c l l c}
\hline
Node $k$ & Demands & Disjoint subset $B_k'$ & Size of $B_k'$ \\
\hline
1 & $\{ W_3^{\{1\}} \}$ 
  & $\{ W_3^{\{1\}} \}$ 
  & $\frac{12T}{14}$ \\

2 & $\{ W_4^{\{2\}},\; W_3^{\{2\}} \}$ 
  & $\{ W_4^{\{2\}},\; W_3^{\{2\}} \}$ 
  & $\frac{12T}{14}$ \\




6 & $\{ W_7^{\{6\}},\; W_8^{\{6\}} \}$ 
  & $\{ W_7^{\{6\}},\; W_8^{\{6\}} \}$ 
  & $\frac{12T}{14}$ \\

7 & $\{ W_8^{\{7\}} \}$ 
  & $\{ W_8^{\{7\}} \}$ 
  & $\frac{12T}{14}$ \\
\hline
\end{tabular}
\caption{Demanded subsegments, corresponding disjoint subsets $B_k'$ and their sizes for the nodes in $N_1' \cup N_2'$, for $K=8,r=6$.}
\label{tab:demands_example2}
\end{table}

As we can see from Fig. \ref{fig:example2a} and \ref{fig:example2b}, the subgraphs are acyclic. Moreover, the sum of the sizes of the subsegments in both the graphs is $\frac{24T}{14}$, which matches the bound given in Theorem \ref{theorem:highr-lowerboundexpression}. It is also easy to see from Table \ref{tab:demands_example2} that all the parts of Lemma \ref{lemma:highrdemandslowerbounds} hold.

Using Lemmas \ref{lemma:highracyclicsubgraphs} and \ref{lemma:highrdemandslowerbounds}, we obtain the following theorem, which is our lower bound in the high-$r$ regime.

\begin{theorem}
\label{theorem:highr-lowerboundexpression}
Let $r\geq\frac{K+1}{2}$. The optimal rebalancing load $\Lrem^*(r)$ for the node removal scenario satisfies the following lower bound, when the target database is cyclic. 
\begin{align}
\label{eqn:lowerbound-highrregime}
    \Lrem^*(r)\geq \max\left(\frac{r(K-r)}{K-1},\frac{r}{r-1}\right).
\end{align}
Further, the following is true. 
\begin{align}
\label{eqn:maxexpression-highr}
    \max\left(\frac{r(K-r)}{K-1},\frac{r}{r-1}\right)=\begin{cases}
    \frac{r(K-r)}{K-1},& \text{if}~K\geq r+2,\\
    \frac{r}{r-1},& \text{if}~K = r+1.
    \end{cases}
\end{align}
\end{theorem}
\begin{IEEEproof}
The proof of the lower bound \eqref{eqn:lowerbound-highrregime} is essentially identical to that of Theorem \ref{theorem:lowr-lowerboundexpression} in combination with the existing lower bound from \cite{krishnan2020coded}. We skip the details to avoid repetition.

We now prove \eqref{eqn:maxexpression-highr}. Let $f(K) = \frac{r(K-r)}{(K-1)}$. Then,
\begin{align*}
f'(K) = \frac{r(r-1)}{(K-1)^2} > 0,
\end{align*}
so $f(K)$ is strictly increasing in $K$. For $K = r+1$, we have $f(r+1) = \frac{r\cdot 1}{r} = 1 < \frac{r}{r-1}$, so the second argument dominates. For $K \geq r+2$, we verify that $f(K) \geq \frac{r}{r-1}$. Indeed, solving $\frac{r(K-r)}{K-1} \geq \frac{r}{r-1}$ gives $K(r-2) \geq r^2-r-1$, which holds for all integers $K \geq r+2$ and $r \geq 3$. Since $K \geq r+1$ and $K$ is an integer, the two cases $K = r+1$ and $K \geq r+2$ are exhaustive, which completes the proof.
\end{IEEEproof}

\subsection{Order-wise Optimality of Achievable Rate in Theorem \ref{thm:achieve} (Proof of Theorem \ref{thm:achieve} (Part \textit{2)})} 
\label{subsec:multiGap_revised}
In this section, we show that the achievable rate in Theorem \ref{thm:achieve} is optimal order-wise, as $K$ and $r$ grow large. Specifically, we show that the achievable rate in Theorem \ref{thm:achieve} is within a multiplicative constant of the lower bounds in Theorems \ref{theorem:lowr-lowerboundexpression} and \ref{theorem:highr-lowerboundexpression}, in the respective regimes of $r,K$, as given in part \textit{2)} of Theorem \ref{thm:achieve}.
\begin{IEEEproof}[Proof of Part \textit{2)} of Theorem \ref{thm:achieve}]
Firstly, we note that, when $r < \frac{K+1}{2}$, the achievable scheme uses Scheme 2. In this case, the load from Theorem \ref{thm:achieve} is
\begin{align}
\label{eqn1}
    \Lrem(r)
    &=
    \frac{K-r}{K-1}
    +
    L_2(r)
    \nonumber \\
    &=
    \frac{1}{2(K-1)}
    \Bigg(
    2(K-r)
    +
    K(r-1)
    +
    \ceil{\frac{r^2-2r}{2}}
    \Bigg).
\end{align}
From Theorem \ref{theorem:lowr-lowerboundexpression}, we have that, in this regime,
\begin{align}
\label{eqn2}
    \Lrem^*(r) \geq \frac{K-r}{K-1}+\frac{K(r-1)}{2(K-1)}. 
\end{align}

Now, in the regime $r<(K+1)/2$, observe from \eqref{eqn1} and \eqref{eqn2} that 
\begin{align*}
\frac{\Lrem(r)-\Lrem^*(r)}
{\Lrem^*(r)}
&\leq
\frac{\frac{(r-1)^2}{2}}
{2(K-r)+K(r-1)} \\
&=
\frac{(r-1)^2}
{4(K-r)+2K(r-1)}.
\end{align*}
Now, we will show that this quantity is $<1/4$. To see that, consider the following
\begin{align*}
&4(K-r)+2K(r-1)-4(r-1)^2 \\
&= 4(K-r)
+(r-1)(2K-4r+4) \\
&\stackrel{(a)}{>}
4(K-r) \\
&\quad +
(r-1)
\left(
2K+4-\frac{4(K+1)}{2}
\right)
>0,
\end{align*}
where in $(a)$ we have used the fact that $r<\frac{(K+1)}{2}$. Thus, we see that 
$\frac{\Lrem(r)-\Lrem^*(r)}{\Lrem^*(r)}<\frac{1}{4}$, and hence $\frac{\Lrem(r)}{\Lrem^*(r)}<\frac{5}{4}$, when $r<\frac{K+1}{2}$.

 Now, we come to the case when $r=\frac{K+1}{2}$. For this case, we have $\Lrem^*(r)\geq \frac{r(K-r)}{K-1}$. Thus, we have the following inequalities.
\begin{align*}
    \frac{\Lrem(r)}{\Lrem^*(r)}&\leq \frac{2(K-r)+K(r-1)+\frac{(r-1)^2}{2}}{2r(K-r)}\\
    &=\frac{4(r-1)+2(2r-1)(r-1)+(r-1)^2}{4r(r-1)}\\
    &=\frac{5r+1}{4r}\\
    &=\frac{5}{4}+\frac{1}{4r}\\
    &\stackrel{(a)}{\leq} \frac{4}{3}.
\end{align*}
where $(a)$ holds since $r\geq 3$.

 Now, we consider the scenario of $\frac{K+1}{2} < r < \min\left(\lceil \frac{2K+2}{3} \rceil, K-1\right)$. Note that this means $r \leq \frac{2K+1}{3}$. In this regime, Scheme 2 gives us the appropriate achievability scheme as per Theorem \ref{thm:achieve}, given in (\ref{eqn1}).
 
From Theorem \ref{theorem:highr-lowerboundexpression}, the lower bound satisfies:
\begin{equation*}
    L_{rem}^*(r) \geq \frac{r(K-r)}{K-1}.
\end{equation*}
We will show that $L_{rem}(r) \leq 2 L_{rem}^*(r)$ in this regime, via the following inequalities:
\begin{align*}
    \frac{L_{rem}(r)}{L_{rem}^*(r)} &\leq \frac{2(K-r) + K(r-1) + \lceil \frac{r^2-2r}{2} \rceil}{2r(K-r)} \nonumber \\
    &\leq \frac{2(K-r) + K(r-1) + \frac{r^2-2r+1}{2}}{2r(K-r)}.
\end{align*}
To prove the ratio is bounded by 2, we must equivalently show that the numerator is bounded by $4r(K-r)$. Taking the difference between the denominator multiplied by 2 and the numerator, we get:
\begin{align*}
    & 4r(K-r) - \left( 2(K-r) + K(r-1) + \frac{r^2-2r+1}{2} \right) \nonumber \\
    &= 4rK - 4r^2 - 2K + 2r - Kr + K - 0.5r^2 + r - 0.5 \nonumber \\
    &= K(3r-1) - 4.5r^2 + 3r - 0.5 \nonumber \\
    &\overset{(a)}{\geq} (1.5r - 0.5)(3r-1) - 4.5r^2 + 3r - 0.5 \nonumber \\
    &= 4.5r^2 - 1.5r - 1.5r + 0.5 - 4.5r^2 + 3r - 0.5 \nonumber \\
    &= 0,
\end{align*}
where (a) holds due to the constraints on $r$ in this regime. Specifically, the regime bound $r \leq \frac{2K+1}{3}$ directly implies $2K \geq 3r-1$, or equivalently $K \geq 1.5r - 0.5$. Because the final evaluation yields $0$, we have strictly established $\frac{L_{rem}(r)}{L_{rem}^*(r)} \leq 2$.

Now, we look at the case when $\lceil \frac{2K+2}{3} \rceil \leq r \leq K-2$. In this scenario, Scheme 1 is applicable, for which the load from Theorem \ref{thm:achieve} is:
\begin{align}
L_{rem}(r)
&=
\frac{
2(K-r)+2(K-r)(2r-1)
}{
2(K-1)
}
\nonumber \\
&=
\frac{4r(K-r)}{2(K-1)}
\nonumber \\
&=
\frac{2r(K-r)}{K-1}.
\label{eq:scheme1_load}
\end{align}
Because $r \leq K-2$, we have $K \geq r+2$, which perfectly satisfies the condition $K \geq r + 1 + \frac{1}{r-2}$ for $r \geq 3$. Using the bound on $L_{rem}^*(r)$ as per the cyclic constraint from Theorem \ref{theorem:highr-lowerboundexpression} applicable here, we have:
\begin{equation*}
    \frac{L_{rem}(r)}{L_{rem}^*(r)} \leq \frac{\frac{2r(K-r)}{K-1}}{\frac{r(K-r)}{K-1}} = 2.
\end{equation*}

Finally, we come to the case when $r = K-1$. In this case, the achievable rate remains as defined by Scheme 1 in \eqref{eq:scheme1_load}. However, because $K = r+1$, the applicable lower bound from Theorem \ref{theorem:highr-lowerboundexpression} strictly defaults to $\frac{r}{r-1}$. Thus, we have:
\begin{equation*}
    \frac{L_{rem}(r)}{L_{rem}^*(r)} \leq \frac{\frac{2r(1)}{r}}{\frac{r}{r-1}} = \frac{2(r-1)}{r} < 2.
\end{equation*}
This completes the proof of part 2) of Theorem \ref{thm:achieve}, confirming that the achievable rate is within a multiplicative factor of 2 for all $r > \frac{K+1}{2}$.

\end{IEEEproof}

\begin{remark}
The multiplicative gap of at most $2$ between our 
achievable rebalancing loads and the lower bounds derived 
in this section bears a structural resemblance to the 
factor-of-$2$ gap established in the context of \textit{embedded} index coding
\cite{porter_embedded}. In the index coding setting assumed in \cite{porter_embedded}, the nodes simultaneously act as senders and receivers (similar to our rebalancing setting) without an additional central server which contains all the data. It is shown 
that the optimal communication load in such a decentralized setting is at most twice than that of an optimal 
centralized solution, i.e., when a central node containing all the data is assumed (Theorem 4 of \cite{porter_embedded}). 
In the present work, the lower bound on the rebalancing load for node-removal was based on the MAIS-based index-coding bound \cite{neely2013dynamic}, which applies to the centralized  (and thus to the decentralized) setting. Thus,
the factor-of-$2$ gap we obtain here, between the lower and upper bounds on the communication load for rebalancing in node-removal (Theorem \ref{thm:achieve})  is consistent with the findings of \cite{porter_embedded}. 
Formalizing this connection more precisely is an interesting 
direction for future work.
\end{remark}

\section{Numerical Comparisons}
\label{sec:lowerboundcomparisons}

\begin{figure}[h]
\includegraphics[scale=0.6]{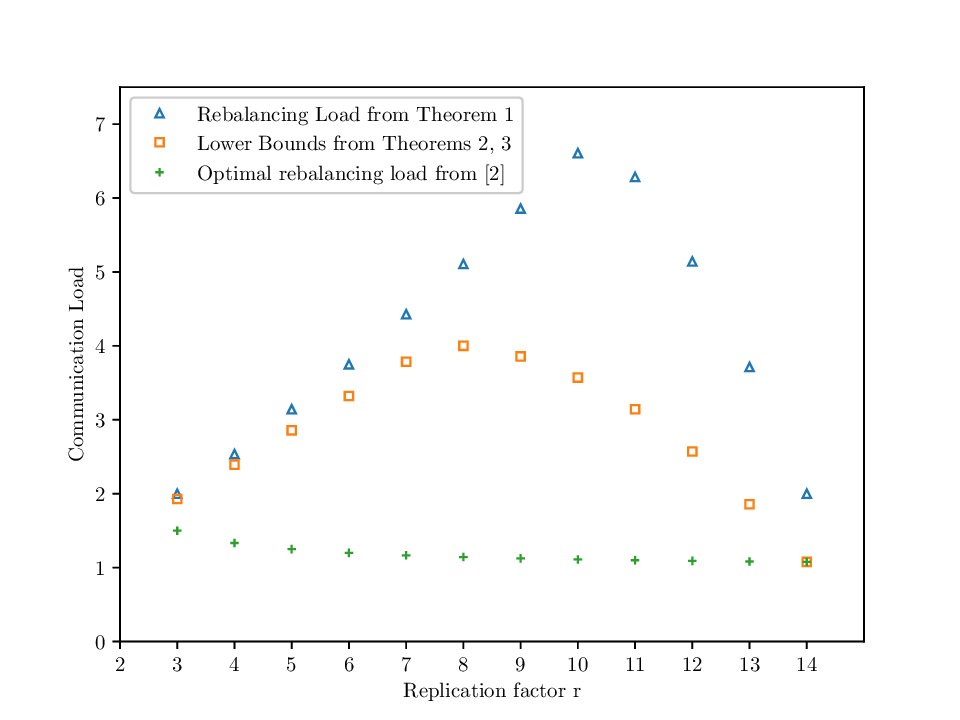}
\centering
\caption{\small Node-removal load comparison: For $K=15$, the figure shows comparisons of the lower bounds from Theorems \ref{theorem:lowr-lowerboundexpression}, \ref{theorem:highr-lowerboundexpression} (cyclic initial and target databases) with the rebalancing load given in Theorem \ref{thm:achieve}, as well as the tight converse (information-theoretic optimal load) achieved by the scheme in \cite{krishnan2020coded} (across all $r$-balanced databases) using a large file size, for varying $r$.}
\label{fig:lowerbound}
\end{figure}

For $K=15$ and $r$ varying in the range $\{3,4,\hdots,14\}$, Fig. \ref{fig:lowerbound} shows a numerical comparison between (a) the lower bounds given in Theorems \ref{theorem:lowr-lowerboundexpression} and \ref{theorem:highr-lowerboundexpression}, (b) the achievable rebalancing load given in Theorem \ref{thm:achieve}, and (c) the tight lower bound based on the results of \cite{krishnan2020coded} ($\mathfrak{L}_{rem}^*(r)N=\frac{rK}{K(r-1)}=\frac{r}{r-1}$). Recall from Algorithm \ref{algo:rebalancing} that Scheme 1 is applicable for $r \geq \ceil{\frac{2K+2}{3}}=11$, otherwise Scheme 2 is applicable. On the other hand, the lower bound from Theorem \ref{theorem:lowr-lowerboundexpression} is applicable for the range $3\leq r\leq (K+1)/2=8$, while Theorem \ref{theorem:highr-lowerboundexpression} is applicable for the range $9\leq r\leq 14$. 


\section{Conclusion}
\label{discussion}
We have presented a XOR-based coded rebalancing scheme for the case of node removal and node addition in $r$-balanced cyclic databases with $K$ nodes. Our scheme only requires the file size to be only cubic in the number of nodes in the system, as compared to the scheme in \cite{krishnan2020coded} which requires exponential file size. Further, the file-size of our schemes was argued to be order-optimal. For the node removal case, we presented a coded rebalancing algorithm that chooses between the better of two coded transmission schemes in order to reduce the communication load incurred in rebalancing. 
 We showed that the communication load of this rebalancing algorithm is always smaller than that of the uncoded scheme. 
For the node addition case, we presented a simple uncoded scheme which achieves the optimal load. For the node-removal case, for the setting where the initial and target databases have cyclic and balanced placement of data, we have presented lower bounds for the load in Theorems \ref{theorem:lowr-lowerboundexpression} and \ref{theorem:highr-lowerboundexpression}. We further showed that the achievable loads are order-optimal within a constant multiplicative factor of this lower bound. In particular, the achievable load is within a multiplicative factor of at most $\frac{5}{4}$, $\frac{4}{3}$, and $2$, from the lower bound for $r < \frac{K+1}{2}$, $r = \frac{K+1}{2}$, and $r > \frac{K+1}{2}$, respectively.


 We conclude with a few remarks on possible future directions.
 \begin{itemize}
    \item Remark \ref{rem:asymptoticgap} indicates that in the regime when both $r$ and $K-r$ grow asymptotically as $\Theta(K)$, the gap between the node-removal rebalancing load on cyclic databases and the optimal load $\frac{r}{r-1}$ (across all $r$-balanced databases) can be $\Theta(K)$, and thus very wide. It would be worthwhile to explore if there are other placement and rebalancing schemes that need small file-size than the optimal-load scheme in \cite{krishnan2020coded}, which asymptotically reduce this wide gap from $\Theta(K)$, in this regime of $r$. 
     \item Designing rebalancing schemes for node-removal with lower communication loads, as well as for other types of placement schemes, using other index coding techniques, such as cycle cover schemes, partial-clique cover schemes, would be interesting. Pliable index coding \cite{song_fragouli_PICOD_datashuffling}, which assumes that each client requests for any message (or a set of $t$ messages) not already in its side information, also seems to be a good candidate for the rebalancing problem. However, if we seek to achieve a specific target database, such as a cyclic database, we then have various constraints on the relationships between node-demands during the rebalancing phase. This makes the rebalancing problem more complex than the scenarios for pliable index coding available in literature, and therefore novel ideas are needed. 
     \item  Finally, to the best of our knowledge, data rebalancing in the context of erasure-coded distributed storage systems remains unexplored, and investigating this setting is a promising avenue for future research.
 \end{itemize}

\section*{Acknowledgements}    
The first two authors would like to acknowledge Shubhransh Singhvi for fruitful discussions on the problem.

\bibliographystyle{IEEEtran}
\bibliography{IEEEabrv,2024_with_lower_bound.bib}

\appendices
\section{Proof of Claim \ref{claim:threshold}}
\label{appendix proof threshold}
First, we can assume $K\geq 4$ without loss of generality, as our replication factor $r$ lies between $\{3,\hdots,K-1\}$.  Consider the expressions in Theorem \ref{thm:achieve} for $L_1(r)$ and $L_2(r)$ as continuous functions of $r$. Also, consider $L_{2,o}(r)=\frac{K-r}{(K-1)}+\frac{r-1}{2(K-1)}\left(K+\frac{r-1}{2}\right)$ be the continuous function of $r$ which matches with $\frac{K-r}{K-1}+L_2(r)$ in Theorem \ref{thm:achieve} for odd values of $r\in\{3,\hdots,K-1\}$. Similarly, let $L_{2,e}(r)=\frac{K-r}{(K-1)}+\frac{r-2}{2(K-1)}\left(K+\frac{r}{2}\right) + \frac{K}{2(K-1)}$ be the continuous function of $r$ which matches with $\frac{K-r}{K-1}+L_2(r)$ in Theorem \ref{thm:achieve} for even values of $r\in\{3,\hdots,K-1\}$. 

Let $r_o$ be a real number in the interval $[3:K-1]$ such that $L_{2,o}(r_o)=L_1(r_o)$. Similarly, let  $r_e$ be a real number $r$ in the interval $[3:K-1]$ such that $L_{2,e}(r_e)=L_1(r_e)$. 

With these quantities set up, the proof then proceeds according to the following steps.
\begin{enumerate}
    \item Firstly, we show that $L_{2,o}(r) > L_{2,e}(r)$ for $K \geq 4$. 
    \item We then find the values of $r_o$ and $r_e$, which turn out to be unique. We also show that $r_e>r_o$ and that $\ceil{r_e}=\ceil{\frac{2K+2}{3}}=\floor{r_o}+1$. 
    \item Then, we shall show that for any integer $r> r_e$, we have $L_1(r)< L_{2,e}(r)$.  Further, we will also show that for any integer $r< r_o$, we have $L_{2,o}(r)<L_1(r)$.  
\end{enumerate}
It then follows from the above steps that the threshold value is precisely $r_{\mathsf{th}}=\ceil{r_e}=\ceil{\frac{2K+2}{3}}$. We now show the above steps one by one. 

 \textit{1) Proof of $L_{2,o}(r) > L_{2,e}(r)$ for $K \geq 4$: } 
 We have that 
\begin{align*}
    L_{2,o}&(r)-L_{2,e}(r)\\
    &=
    \frac{r-1}{2(K-1)}
    \left(
    K+\frac{r-1}{2}
    \right)
    \nonumber \\
    &\quad -
    \frac{r-2}{2(K-1)}
    \left(
    K+\frac{r}{2}
    \right)
    -
    \frac{K}{2(K-1)} \\
    &=
    \frac{
    (r-1)(2K+r-1)
    -(r-2)(2K+r)
    -2K
    }{
    4(K-1)
    } \\
    &=
    \frac{
    2rK-2K+r^2-2r+1
    }{
    4(K-1)
    }
    \nonumber \\
    &\quad -
    \frac{
    2rK-4K+r^2-2r+2K
    }{
    4(K-1)
    } \\
    &=
    \frac{1}{4(K-1)}
    >0,
\end{align*}

which holds as $K\geq 4$. Hence, $L_{2,o}(r) > L_{2,e}(r)$ for $K \geq 4$.


\textit{2) Finding $r_o$ and $r_e$ and their relationship: } Calculating $L_{2,o}(r)-L_1(r)$, we get the following
\begin{align*}
    L_{2,o}(r) - L_1(r) &=\frac{r-1}{2(K-1)}\left(K+\frac{r-1}{2}\right) - \frac{(K-r)(2r-1)}{(K-1)}\\
    &= \frac{(r-1)(2K+r-1) - 4(K-r)(2r-1)}{4(K-1)}  \\
    &= \frac{9r^2 - 6r(K+1) + 2K+1}{4(K-1)}.
\end{align*}
Solving for $r$ from $L_{2,o}(r)-L_1(r)=0$ with the condition that $r\geq 3$, we get 
\begin{align}
\label{eqn:r_intodd}
    r&=\frac{2K+1}{3}.
    \nonumber
\end{align} 
It is easy to see that $\frac{2K+1}{3}>2$ for $K\geq 4$. Also, we can check that $\frac{2K+1}{3}< (K-1)$, when $K\geq 4$. Thus, we have that $r_o=\frac{2K+1}{3}$. By similar calculations for $L_{2,e}(r)-L_1(r)$, we see that $r_e=\frac{K+1 + \sqrt{K^2+1}}{3}$.
Further, we observe that $r_e - r_o = \frac{\sqrt{K^2+1}-K}{3}$, for $K \geq 4$. Hence, $r_e > r_o$ for $K \geq 4$. Also observe that $\ceil{r_e}\leq \ceil{\frac{2K+2}{3}}=\floor{\frac{2K+1}{3}}+1=\floor{r_o}+1$, where the first equality holds as $K\geq 4$. Now, as $r_e>r_o$ we have that $\ceil{r_e}>\floor{r_o}$. Thus, we see that $\ceil{r_e}=\ceil{\frac{2K+2}{3}}=\floor{r_o}+1$.

    \textit{Proof of 3):} We now prove that for any integer  $r$ if $r> r_e$, then $L_1(r) < L_{2,e}(r)$. Let $r = r_e + a$, for some real number $a > 0$ such that $r_e + a$ is an integer. We know that $L_1(r_e) = L_{2,e}(r_e)$. Consider the following sequence of equations. 
    \begin{align*}
        L_1(r)
        &= L_1(r_e+a) \\
        &=
        \frac{K-(r_e+a)}{K-1}
        +
        \frac{(K-(r_e+a))(2(r_e+a)-1)}
        {K-1} \\
        &=
        \frac{K-r_e}{K-1}
        -
        \frac{a}{K-1}
        \nonumber \\
        &\quad +
        \frac{
        (K-r_e-a)(2r_e-1+2a)
        }{
        K-1
        } \\
        &=
        \frac{K-r_e}{K-1}
        +
        \frac{(K-r_e)(2r_e-1)}
        {K-1}
        \nonumber \\
        &\quad +
        \frac{
        -a-a(2r_e-1+2a)+2a(K-r_e)
        }{
        K-1
        } \\
        &=
        L_1(r_e)
        +
        \frac{2aK-2a^2-4ar_e}{K-1} \\
        &=
        L_{2,e}(r_e)
        +
        \frac{2aK-2a^2-4ar_e}{K-1}.
\end{align*}
    Similarly, consider the following.
    \begin{align*}
        L_{2,e}(r)
        &= L_{2,e}(r_e+a) \\
        &=
        \frac{K-(r_e+a)}{K-1}
        +
        \frac{(r_e+a)-2}{2(K-1)}
        \left(
        K+\frac{r_e+a}{2}
        \right) \\
        &=
        \frac{K-r_e}{K-1}
        -
        \frac{a}{K-1}
        \nonumber \\
        &\quad +
        \frac{(r_e-2)+a}{2(K-1)}
        \left(
        K+\frac{r_e+a}{2}
        \right) \\
        &=
        \frac{K-r_e}{K-1}
        +
        \frac{
        (r_e-2)\left(K+\frac{r_e}{2}\right)
        }{
        2(K-1)
        }
        \nonumber \\
        &\quad +
        \frac{
        -6a+a^2+2aK+2ar_e
        }{
        4(K-1)
        } \\
        &=
        L_{2,e}(r_e)
        +
        \frac{
        -6a+a^2+2aK+2ar_e
        }{
        4(K-1)
        }.
\end{align*}
    \begin{align*}
        \text{Now, }L_{2,e}(r) - L_1(r) &= \frac{18ar_e - 6aK + 9a^2 -6a}{4(K-1)} \\
        &= \frac{a(18r_e-6K+9a-6)}{4(K-1)} \\
         &\stackrel{(a)}> \frac{a(6\sqrt{K^2+1}+9a)}{4(K-1)} \\
         &\stackrel{(b)}> 0,
    \end{align*}
    where (a) holds on substituting $r_e = \frac{K+1+\sqrt{K^2+1}}{3}$,  (b) holds as $K, a >0$.  Therefore, when $r> r_e$, $L_1(r)<L_{2,e}(r)$.
    
We now prove that for any integer  $r$ if $r< r_o$, then $L_{2,o}(r)<L_1(r)$. Let $r = r_o - a$, for some real number $a < r_o$ such that $r_o - a$ is an integer. We know that $L_1(r_o) = L_{2,o}(r_o).$ Consider the following sequence of equations.
    \begin{align*}
        L_1(r)
        &= L_1(r_o-a) \\
        &=
        \frac{K-(r_o-a)}{K-1}
        +
        \frac{(K-(r_o-a))(2(r_o-a)-1)}
        {K-1} \\
        &=
        \frac{K-r_o}{K-1}
        +
        \frac{a}{K-1}
        \nonumber \\
        &\quad +
        \frac{
        (K-r_o+a)(2r_o-1-2a)
        }{
        K-1
        } \\
        &=
        \frac{K-r_o}{K-1}
        +
        \frac{(K-r_o)(2r_o-1)}
        {K-1}
        \nonumber \\
        &\quad +
        \frac{
        a+a(2r_o-1-2a)-2a(K-r_o)
        }{
        K-1
        } \\
        &=
        L_1(r_o)
        +
        \frac{4ar_o-2aK-2a^2}{K-1} \\
        &=
        L_{2,o}(r_o)
        +
        \frac{4ar_o-2aK-2a^2}{K-1}.
\end{align*}
    Similarly, consider the following.
    \begin{align*}
        L_{2,o}(r)
        &= L_{2,o}(r_o-a) \\
        &=
        \frac{K-(r_o-a)}{K-1}
        +
        \frac{(r_o-a)-1}{2(K-1)}
        \left(
        K+\frac{(r_o-a)-1}{2}
        \right) \\
        &=
        \frac{K-r_o}{K-1}
        +
        \frac{a}{K-1}
        \nonumber \\
        &\quad +
        \frac{(r_o-1)-a}{2(K-1)}
        \left(
        K+\frac{(r_o-1)-a}{2}
        \right) \\
        &=
        \frac{K-r_o}{K-1}
        +
        \frac{
        (r_o-1)\left(
        K+\frac{r_o-1}{2}
        \right)
        }{
        2(K-1)
        }
        \nonumber \\
        &\quad +
        \frac{
        6a-2ar_o+a^2-2aK
        }{
        4(K-1)
        } \\
        &=
        L_{2,o}(r_o)
        +
        \frac{
        6a-2ar_o+a^2-2aK
        }{
        4(K-1)
        }.
\end{align*}
    \begin{align*}
        \text{Now, }L_1(r) - L_{2,o}(r) &= \frac{18ar_o - 6aK - 9a^2 -6a}{4(K-1)} \\
        &= \frac{a(18r_o - 6K - 9a -6)}{4(K-1)} \\
        &= \frac{3a(3r_o-3a+3r_o-2K-2)}{4(K-1)} \\
        &\stackrel{(a)}\geq \frac{3a(3+3r_o-2K-2)}{4(K-1)} \\
        &\geq \frac{3a(3r_o-2K+1)}{4(K-1)} \\
        &\stackrel{(b)}\geq \frac{3a(1+1)}{4(K-1)}  \\
        &> 0,
        \end{align*}
    where (a) holds because $r_o-a \geq 1$ and (b) holds on substituting $r_o=\frac{2K+1}{3}$. Therefore, when $r< r_o$, $L_{2,o}(r)<L_1(r)$. The proof is now complete.

\section{Computational Complexity of the Rebalancing Algorithm for Node-Removal}
\label{app:computationcomplexitynoderemoval}
We analyze the computational complexity of the rebalancing algorithm for node-removal by breaking it down into different phases namely splitting, transmission, decoding, and merging, as described in Section \ref{sec:rebalancing_noderemoval}. Let us recall that there are total $K$ segments initially, each of size $T$ bits and there are total $r$ segments in the removed node.

\begin{itemize}
    \item[1.] Splitting Phase: From lines 3, 7, 8, 11, 12 of Algorithm \ref{algo:split}, it is easy to see that each of the segments from the removed nodes is split into up to $O(K)$ subsegments, but every bit of the $r$ segments in the removed node is processed  only once during the process. Therefore, the runtime of the splitting phase is $O(rT)$.   
    \item[2.] Transmission Phase: From lines 6, 7 of Algorithms \ref{algo:case1} and \ref{algo:case2}, the total number of bits in the uncoded broadcasts in each of these algorithms is $O(2T)$. However, the number of XOR-coded broadcasts is $O(K)$  for Scheme 1 (see line 2 of Algorithm \ref{algo:case1}) and $O(r)$ for Scheme 2 (see line 2 of Algorithm \ref{algo:case2}). Moreover, the number of subsegments in the XOR-coded broadcasts is $O(r)$ for Scheme 1 (see line 3, 4 of Algorithm \ref{algo:case1})  and $O(1)$ for Scheme 2 (see line 3, 5 of Algorithm \ref{algo:case2}). We know that the XOR of two bit-strings of length $L$ costs $\Theta(L)$ bit operations. Here, the length of each subsegment is $O(T)$. Therefore, the total number of operations for Scheme 1 is $O(KrT)$ and for Scheme 2 is $O(rT)$. 
    \item[3.] Decoding Phase: Once the broadcasts are received by the nodes, they either XOR-out the subsegments that they already have in case of XOR-coded broadcasts or simply copy the subsegment if it is uncoded. For uncoded broadcasts, the operation is simply copying the subsegment and we know that copying a bit-string of length $L$ takes $\Theta(L)$ operations. Therefore, the total cost involved in copying the uncoded segments is $\Theta(T)$. For XOR-coded broadcasts, if the number of subsegments in the broadcast is $S$, then $S-1$ XOR operations of bit-strings of length $O(T)$ have to be performed to recover a subsegment, at any node that desires the same. Now, as per the superscripts assigned in the splitting phase which indicate the target nodes for any subsegment, we see that every subsegment is to be delivered to at most $r$ nodes. Therefore, the total number of operations for Scheme 1 is $O(r.KrT)=O(Kr^2T)$ and for Scheme 2 is $O(r^2T)$. 
    \item[4.] Merging Phase: From lines 3, 7, 10, 14, 17, 19 of Algorithm \ref{algo:merge}, we can see that each of the survivor nodes concatenates about $O(T)$ bits per segment, with the bits it already had. Therefore, the total number of operations is $O(KT)$.
\end{itemize}
Putting these together, the overall time complexity depends on which scheme is used. For large $r$, Scheme 1 dominates, and the runtime is about $O(Kr^2T)$, which is at most $O(K^3T)$. For small $r$, Scheme 2 dominates, and the runtime reduces to about $O(K^2T)$. Thus, the algorithm always runs in polynomial time, scaling linearly with the segment size $T$ and at most cubical in the number of nodes $K$.

\section{Computational Complexity of the Rebalancing Algorithm for Node-Addition}
\label{app:computationcomplexitynodeaddition}
We analyze the computational complexity of the rebalancing algorithm for node-addition (Algorithm \ref{algo:nodeaddition}).

From lines 2--3, every bit of each of the $K$ segments 
is processed during splitting, resulting in a runtime 
of $O(KT)$. Each node then broadcasts the subsegment 
$W_i^{\{K+1\}\cup[\min(r-1,i-1)]}$ of size $\frac{T}{K+1}$ 
(see line 4), contributing $O(T)$ to the overall runtime. 
Thus the dominant cost of lines 2--5 is $O(KT)$. 

From lines 7-8, total number of bits processed by each of the $r$ nodes is $\frac{KT}{K+1} = O(T)$, so across $r$ nodes it is $O(rT)$. Next, it is easy to see  that operations in lines 10-11 contribute $O(rT)$ to the runtime. Finally, discarding of bits in lines 13-14 add up to $O(rT)$ to the runtime.

Putting all these together, the dominant operation is splitting with complexity $O(KT)$, thus making the overall complexity linear in the number of segments $K$ and linear in bits-per-segment $T$.

\end{document}